\documentclass[10pt, conference, compsocconf]{IEEEtran}
\usepackage{amsmath}
\usepackage{graphicx}
\usepackage{makeidx}  
\usepackage{graphics}
\usepackage{subfigure}
\usepackage{epstopdf}
\ifCLASSINFOpdf

\else

\fi

\begin{document}

\title{Using an Artificial Financial Market for studying a Cryptocurrency Market}
\author{\IEEEauthorblockN{Luisanna Cocco, Giulio Concas and Michele Marchesi}
\IEEEauthorblockA{University of Cagliari, Italy\\
Dipartimento Ingegneria Elettrica ed Elettronica\\
Email: [luisanna.cocco, concas, michele]@diee.unica.it}
}
\maketitle

\begin{abstract}
This paper presents an agent-based artificial cryptocurrency market in which heterogeneous agents buy or sell cryptocurrencies, in particular Bitcoins. In this market, there are two typologies of agents, Random Traders and Chartists, which interact  with each other by trading Bitcoins.
Each agent is initially endowed with a finite amount of crypto and/or fiat cash and issues buy and sell orders, according to her strategy and resources.
The number of Bitcoins increases over time with a rate proportional to the real one, even if the mining process is not explicitly modelled.

The model proposed is able to reproduce some of the real statistical properties of the price absolute returns observed in the Bitcoin real market. 
In particular, it is able to reproduce the autocorrelation of the absolute returns, and their cumulative distribution function.
The simulator has been implemented using object-oriented technology,  and could be considered a valid starting point to study and analyse the cryptocurrency market and its future evolutions.
\end{abstract}

\begin{IEEEkeywords}
Artificial Financial Market, Cryptocurrency, Bitcoin, Heterogeneous Agents, Market Simulation
\end{IEEEkeywords}

\IEEEpeerreviewmaketitle

\section{Introduction}\label{intro}

Cryptocurrencies are digital currencies alternative to the legal ones.
A cryptocurrency is a computer currency whose implementation is based on the principles of cryptography, used both to validate the transactions and to generate new currency. 
The cryptocurrency implementation often use a proof-of-work scheme recording all transactions in a public ledger in order to protect sellers from fraud.
Most of cryptocurrencies are designed to gradually introduce new currency, placing a ceiling on the total amount of money in circulation, to avoid the inflation phenomena as often happens for "fiat" currencies.

The most popular cryptocurrency is undoubtedly Bitcoin. 
It was created by a computer scientist known as "Satoshi Nakamoto" whose real identity is still unknown \cite{Satoshi}.
Like the other cryptocurrencies, Bitcoins use cryptographic techniques, and thanks to an open source system anyone is allowed to control and modify the source code of the Bitcoin software.
The Bitcoin network is a peer-to-peer network that checks and monitors both the generation of new Bitcoins, (aka "mining") and the transactions in Bitcoins. This network includes a high number of computers connected to each other through the Internet. It performs complex mathematical procedures which give life to the mining and verify the correctness and truthfulness of the Bitcoin transactions.

The Bitcoin system provides a ceiling on the amount of money in circulation, equal to 21 million of Bitcoins, consequently there is not the risk that the number of coins increases too much, devaluating the currency.

Bitcoin has several attractive properties for consumers.
At first, it does not rely on a central bank or a government to regulate the money supply.
It enables quasi- anonymous transactions, providing a greater anonymity than traditional electronic payments.
In addition, Bitcoin transactions are irreversible and can also be very small. Indeed, a Bitcoin transaction can involve only one "Satoshi", a subunit equal to $10^{-8}$ of a Bitcoin.

The Bitcoin can be purchased on appropriate websites such as Crypto Trade CoinMKT, BTC-and Vircurex, Cryptsy, Coinbase, UpBit and Vault of Satoshi, that allow to change fiat cash in Bitcoins. Other sites offer online services, or goods exchange for goods, and accept payments in Bitcoins. The Bitcoin allow everyone to send cryptocurrency internationally at a very small expense.

Over the past years, interest in digital currencies has increased.
Indeed, Bitcoin had a rapid growth, both in value and in the number of transactions since its beginning in early 2009. 
The \textit{BlockChain} \footnote{BlockChain is a web site which displays detailed information about all transactions and Bitcoin blocks, providing graphs and statistics on different data, (https://blockchain.info/). We used this web site for extracting the empirical data (such as daily data about price, unique address number and bitcoin number, in the period between January 1, 2012 and April 10, 2014) used in this work.} Web site provides different graphs and statistical analysis about Bitcoins. In particular, we can observe the time trend of the Bitcoin price.


Between January 2009 and January 2010	there were no exchanges on the market.
Between February 2010 and May 2010 two consumers made the first real-world transactions. One bought 2 pizzas for 10,000 BTC, and another auctioned 10,000 BTC for \$50.

In June 2010, the price grew from \$0.008 to \$0.08 for 1 bitcoin. Thereafter, the price slowly rose until a peak of \$1,150 was reached in December 2013.
In the same month, the Bitcoin price crashed to \$600, rebounded to \$1,000, then crashed again to the \$500 range. 
In January 2014 the price settled in the \$800-\$900 range and in
February and March it fell following the shutdown of historical MTGOX exchange site
and reports regarding Bitcoin ban in China. As of April 2014, one Bitcoin is priced at about \$400. 

The recent attention given to Bitcoin and in general to the cryptocurrencies shows that many consumers are turning their attention toward new trading means to simplify their financial lives. Online purchases performed with cryptocurrency are anonymous, faster and simpler than the traditional credit cards ones.

While the popularity of cryptocurrencis has grown quickly, they still face important argument because of their unconventional way of working.

A lively debate is ongoing about  promise, perils and risks of digital currencies and in particular of Bitcoin. Several papers appeared on these topics, but the attempts to study the cryptocurrency market as a whole are very few.
In this work, we propose an agent-based model aimed to study and analyse the Bitcoin market as a whole. We try to reproduce the main stylized facts present in the real Bitcoin market, such as the autocorrelation and the cumulative distribution function of the price absolute returns.
The model proposed simulates the Bitcoin transactions, by implementing a mechanism for the formation of the Bitcoin price, and a specific behavior for each typology of trader. The paper is organized in the following. In Section \ref{sec:1} we discuss other works related to this paper, in Section \ref{sec:2} we present our model in detail; Section \ref{sec:6} deals with the calibration of the model,
and with the values given to several parameters of the model. Section \ref{sec:7} presents the results of the simulations, including an analysis of Bitcoin real prices. The conclusions of the paper are reported in Section \ref{sec:8}.

\section{Related Work}\label{sec:1}
The study and analysis of the cryptocurrency market is a relatively new field. 
In the last years, several papers appeared on this topic given its potential interest, and the many issues related to it.

Androulaki et al. \cite{Androulaki} studied the privacy guarantees of Bitcoin when Bitcoin is used as a primary currency for the daily transactions. Moore \cite{Moore}, Hout et al. \cite{Hout}, Eyal et al. \cite{Eyal}, Brezo et al. \cite{Brezo} and Hanley \cite{Hanley} analysed promise, perils, risks and issues of digital currencies.
Bergstra et al. \cite{Bergstra} investigated technical issues about protocols and security, and issues about legal, ethical and phychological aspects of cryptocurrencies.
Singh et al. \cite{Singh} proposed an additional layer of mutual trust between the payer and the paye,e in order to enhance the security associated with fast transactions for the real Bitcoin transaction network. 


Only very few attempts have been made so far in order to model the cryptocurrency market as a whole.
Luther \cite{Luther} studied why cryptocurrencies failed to gain widespread acceptance using a simple agent model. The proposed a model  in which crypto-anarchists, computer gamers, tech savy and black market agents derive a specific utility by using the fiat currency or the crypto currency. The utility's value varies with the typology of currency and traders. It takes into account the network related benefits from using  the same money as other agents, the benefits unrelated to networks, and the switching costs that incur to switch to the alternative currency.
The author showed that cryptocurrencies like Bitcoins cannot generate widespread acceptance in absence of significant monetary instability, or of government support, because of the high switching costs and of the importance of the network effects. 

Considering that hundreds of cryptocurrencies have be already proposed on the Internet, and that most of them are waiting for acceptance, Bornholdt and Sneppen \cite{Bornholdt} proposed a model based on Moran process to study the cryptocurrencies able to emerge. This model simulate the interchange between several markets where different cryptocurrencies are traded. In particular, the authors simulate the agent trading, and the mining of new coins at a constant rate, and the communications among traders. They showed that all the crypto-currencies are innately interchangeable, and that the Bitcoin currency in itself is
not special, but may rather be understood as the contemporary dominating crypto-currency that
might be replaced by other currencies.

Since very few works have been made to model the crypto market, in this paper we propose a complex agent-based model to study the  cryptocurrencies market as a whole and to reproduce the main stylized facts present in this market, such as autocorrelation and distribution function of the price absolute returns. 
Our model is inspired by artificial financial market models. These models  are stylized heterogeneous agent models (HAMs) and reproduce the real functioning of markets, trying to explain the main stylised facts observed in financial markets, such as the fat-tailed distribution of returns, the volatility clustering, and the unit-root property. Many works have been published on this topic.

LeBaron \cite{LeBaron} offered a first review of the work appeared in this field.
More recently, Chakraborti et al. \cite{Chakra}, and Chen et al. \cite{Chen} offered other, updated reviews.

Palmer et al. \cite{Palmer} and Arthur et al. \cite{Arthur} proposed an artificial markets combining market trading mechanisms with inductive agent learning. 
In the 2000s, researchers at the University of Genoa and
Cagliari developed the Genoa Artificial Stock Market (GASM). 
In particular, Raberto et al. \cite{Raberto2001} proposed an agent-based artificial financial market that, through a realistic trading mechanism for price formation, is able to reproduce some of the main stylised facts observed in real financial markets.
Raberto et al. \cite{Raberto2002}, and Cincotti et al. \cite{Cincotti}
studied the long-run wealth of traders
characterized by different trading strategies.
Moreover, Raberto et al \cite{Raberto2005} presented an extension of the GASM including a limit order book mechanism for price formation. They demonstrated that the main stylized facts in financial market can be reproduced as a consequence of the limit order book, not introducing any assumption on agents behavior.
Alfarano et al. \cite{Alfarano2008}, starting from the study of relatively complicated agent-based models which do not allow for analytical solutions, proposed a closed-form model that gives rise to realistic behavior of the resulting time series, like fat tails of returns and temporal dependence of volatility.

Liua et al. \cite{Liua2008} developed two simple
models to investigate important statistical features of stock price series. With the first model, the authors found that the clearing house microstructure can explain fat tail, excess volatility and autocorrelation phenomena of high-frequency returns. With the second model the authors investigated the effects of agents' behavioral assumptions on daily returns.
Ponta et al. \cite{Ponta2011} studied the statistical properties of prices and returns by using a heterogeneous agent model.
They simulate an artificial stock market where agents are modelled as nodes of sparsely connected graphs. The agents own an amount of cash and stocks, share information by means of interactions
determined by graphs and trade risky assets; whereas a central market maker determines
the price at the intersection of the demand and supply curves.
Ponta et al. \cite{Ponta2012} proposed a heterogeneous agent model for the simulation of high-frequency market data by using the Genoa Artificial Stock Market. In this market, agents have zero
intelligence and trade a risky asset, the price being cleared by means of a limit order book in which the waiting-time distribution between consecutive orders follows
a Weibull distribution. The authors demonstrated that this mechanism
is able to reproduce fat-tailed distributions
of returns without ad-hoc behavioral assumptions on agents. 

Recently, Feng et al. \cite{Feng} combine the agent-based approach with the stochastic process
approach and propose a model based on the empirically proven behavior of individual market
participants that quantitatively reproduces fat-tailed return distributions and
long-term memory properties. Westerhoff and Franke \cite{Westerhoff} propose a model using three groups of traders: chartists,
fundamentalists and investors, demonstrateing that this combination, together with a simple asset pricing model, can contribute
to explaining the stylized facts of the daily returns of financial markets.

\section{The Model }\label{sec:2}

The proposed model presents an agent-based artificial cryptocurrency market in which heterogeneous agents buy or sell cryptocurrency. In particular, we used the Bitcoin market as a reference to calibrate the model and to compare the results. For the same reason, the fiat currency is referred as "dollars", or "\$".

The features of the model that deserve special mention are:
\begin{itemize}
 \item the trading mechanism is based on a realistic order book that keeps sorted lists of buy and sell orders, and matches them allowing to fullfill compatible orders and to set the price;
 \item traders have typically limited financial resources, initially distributed following a power law;
 \item the number of agents engaged in trading at each moment is a fraction of the total number of agents;
 \item either a number of new traders, endowed only with cash, enter the market every day, or a number of traders quit any trading activity; they represent people who decided to start trading in Bitcoins, or people who decides to quit trading in Bitcoins;
 \item to account for the mining of new Bitcoins, from time to time some traders are randomly chosen and their Bitcoin amount is increased.
 \end{itemize} 
 
The trading mechanism gives rise to a demand-supply schedule, whose imbalance between demand and supply causes price fluctuations.
The limited financial resources of the agents pose significant constraints on the trading of each agent. The engagement in trading of a small fraction of population makes the trading mechanism realistic, given also the fact that the Bitcoin exchange mechanism do not allow high-frequency trading.

The entering the market of new traders interested in buying Bitcoins is an empirical fact, that is reflected in the increasing number of IP addresses registered in the Blockchain. This continuous inflow of traders fits empirical data, and makes the whole market open. These new traders are endowed only of cash, because they represent newcomers to the Bitcoin market, wishing to buy Bitcoins for the first time.
We also account for the ever increasing number of Bitcoins due to mining by increasing proportionally the amount of Bitcoins owned by randomly selected traders who already own Bitcoins, and who act as "miners" in the model.

As regards trading strategies, it is worth noting that traders operate in the market either for real needs, or for speculative reasons. In fact, it is misleading to claim that the agents trade only for speculative reasons, so in the proposed model two populations of traders are defined. One population issues orders for real needs and the other for speculative reasons, hence each of them follows specific trading strategies. 

In the next subsections we described in detail the model, which simulates the Bitcoin transactions and the related mechanism of price formation.

\subsection{The Traders}\label{sec:5}
At every generic time step, $t$, \textit{i}-th trader holds and amount $c_i(t)$ of fiat currency (cash, in dollars), and an amount $b_i(t)$ of crypto-currecy (Bitcoins). 
Traders are divided into "historic" ones - who are in the market since its beginning and own both cash and Bitcoins - and new traders - who enter the market in a second time and are endowed only with cash. 
The wealth distribution of both kinds of traders follows an inverse power-law. This is a realistic assumption because the financial "power" of traders hugely varies. More detailed explanations of this property can be found in \cite{Levy}.

The set of traders entering the market at times $> 0$ are generated in the beginning of the simulation with a Pareto distribution of fiat cash, and then are randomly extracted from the set, when a given number of them must enter the market at a given time step.
More details on how wealth is actually given to traders is given in Section \ref{sec:6.1}.

At each time $t$, an active trader can issue only one order, which can be a sell order or a buy order. Orders already placed but not yet satisfied or withdrawn are accounted for when determining the amount of Bitcoins a trader can buy or sell.
A buy order implies that a trader buys Bitcoin currency in exchange of a given amount of dollars. Consequenty, the amount of Bitcoins to be bought is given by the ratio between the amount of dollars to be traded and the Bitcoin unitary price. 

As regard the trading strategy, traders are divided into two populations: Random traders and Chartists.

\begin{enumerate}
\item \textbf{Random Traders} represent persons who entered the crypto-currency market for various reasons, but not for speculative purposes. They issue orders for reasons linked to their needs, for instance they invest in Bitcoins to diversify their portfolio, or they disinvest to satisfy a need for cash. In the model, this is represented by the fact that they issue orders in a random way, compatibly with their available resources. Buy or a sell orders are always issued by these agents with the same probability.
Random traders' behavior gives stability to the system. They can be considered as a "thermal bath" where other strategies can be introduced.
\item \textbf{Chartists} represent speculators, aimed to gain by placing orders in the Bitcoin market. They speculates that, if prices are rising, they
will keep rising, and if prices are falling, they will keep falling. To this purpose, they issue orders when the price relative variation in a given time window $\tau_i$ is higher than a given threshold $Th$. 

Chartists usually issue buy orders when the price is increasing and sell orders when the price is decreasing. 
Chartists' behavior is key to produce large price variations, and to the reproduction of the basic statistical proprieties of the real returns.
\end{enumerate}

At each time step $t$ only a fraction of traders is active, and hence enabled to issue orders. Random Traders  are active with a probability $p^t_R = 0.1$, whereas the Chartists are active with a probability  $p^t_C = 0.5$. This because the interest of  Chartists in purchasing or selling Bitcoins is higher than that of Random traders. Random traders issue orders to satisfy their needs, whereas Chartists issue order for speculative reasons, study carefully the price variation over time and are more ready to place orders.

The number of traders may vary with time. New traders may enter the market, and existing traders may quit. We assumed that each unique address in the Blockchain corresponds to a trader, and that the number of unique addresses that place orders in a given day is proportional to the number of active traders. To make these data less volatile, we computed and used as an estimate the moving average of the number of unique adresses over a period of 30 days. 

\begin{figure}[!ht]
\centering
\subfigure[]{
\includegraphics[width=0.45\textwidth]
{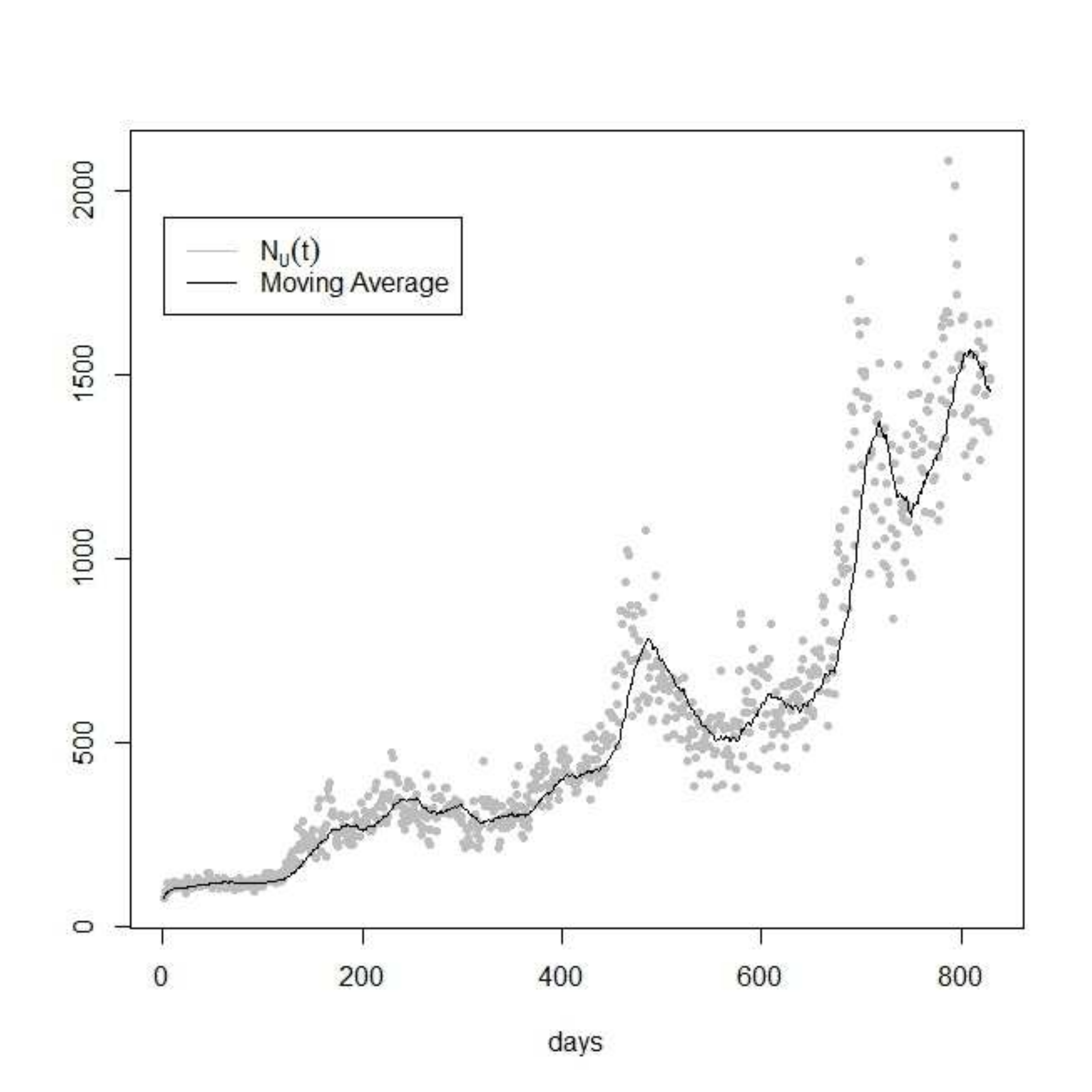}}
\hspace{7mm}
\subfigure[]{
\includegraphics[width=0.45\textwidth]{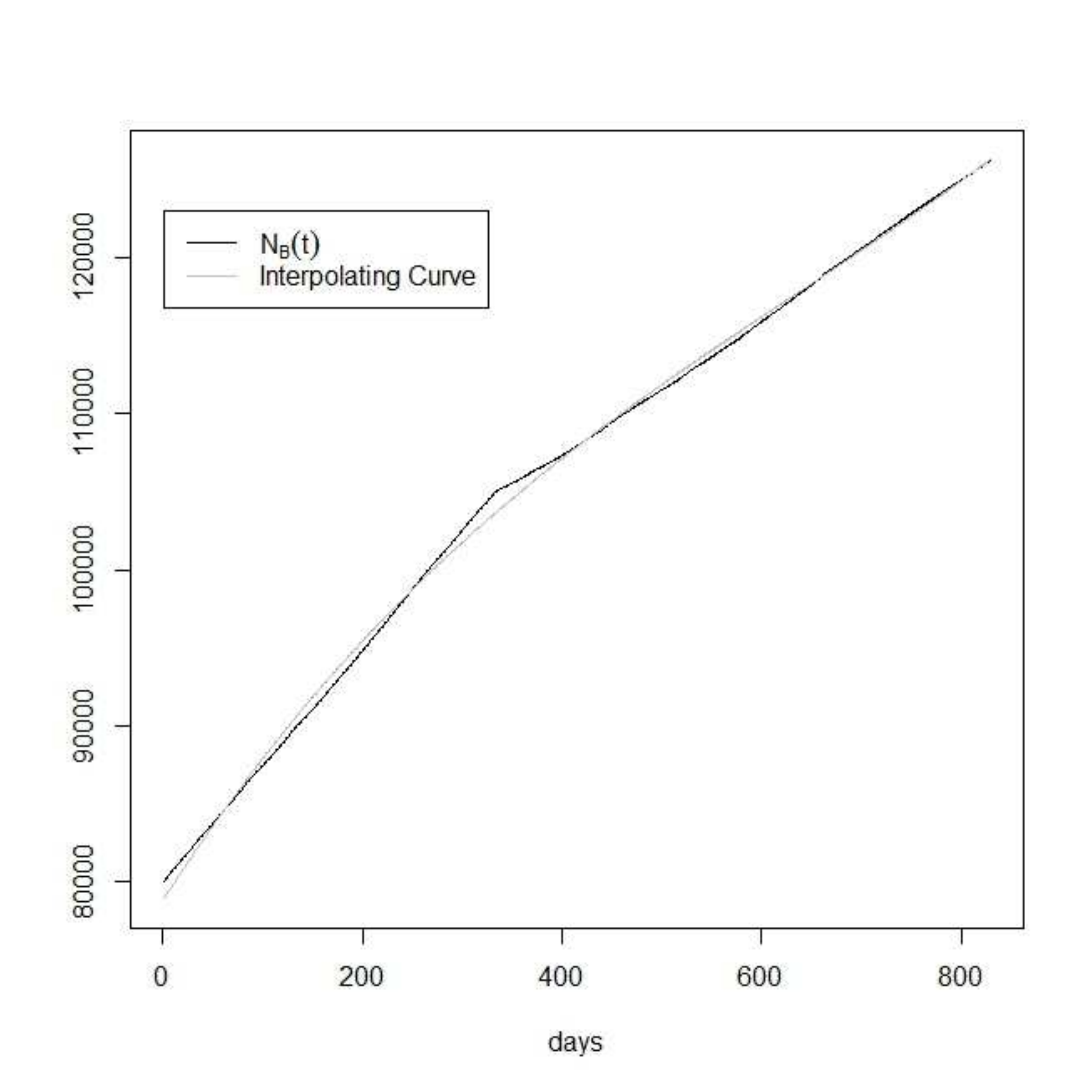}}
\caption{ (a) Number of unique addresses in the Blockchain and their moving average, and  (b) Total number of Bitcoins and its interpolation, in the considered time interval of 830 days.} \label{fig:basicData}
\end{figure}

The average number of traders can increase or decrease with time. If the number increases, we randomly select the proper number of traders from a set of waiting traders, and make them enter the market. If the number decreases, a proper number of traders is randomly selected and removed from the market. Fig. \ref{fig:basicData} (a) shows the number of unique addresses in the considered period, together with the moving average we used.
When a trader is removed, we assume that she does not want to trade Bitcoins anymore, so she posts a market order to sell all her
Bitcoins, and then she is removed from the market. 

At the initial time, the sum of the Bitcoins owned by each trader represents the total number of Bitcoins in the market. 
Over time, Bitcoins are added due to the mining of new Bitcoins. The input rate of new Bitcoins is proportional to the difference between the known total number of Bitcoins in two consecutive days.

We computed this rate and fitted a $4-th$ degree polynomial through it, using as data source the Web site \textit{BlockChain} in the period between January 1, 2012 and April 10, 2014.
Its expression is given by:

\begin{equation}\label{newBitcoin}
N_{B}(t)=4.709*10^{-5}*t^3-0.08932*t^2+98.88*t+78,880
\end{equation}

Fig. \ref{fig:basicData} (b) shows the total number of Bitcoins and their interpolation curve $N_B(t)$.

Since the input rate of bitcoins is always positive, the new Bitcoins which enter the market are assigned to a randomly extracted fraction of Random Traders, giving them an amount of Bitcoins proportional to the Bitcoins already owned. 
The number of traders extracted is equal to the number of new Bitcoins which enter the market, so that on average each selected trader gets one Bitcoin.

\subsection{Buy and Sell Orders}\label{sec:3}
The Bitcoin market is modelled as a steady inflow of buy and sell orders, placed by the traders.
Both buy and sell orders are expressed in Bitcoins, that is, they refer to a given amount of Bitcoins to buy or sell. 
In deeper detail, all orders have the following features:
\begin{itemize}
\item amount, expressed in Bitcoins: it is a real number, because Bitcoins can be bought and sold in fractions as small as a "Satoshi" ($10^{-8}$ of a Bitcoin);
\item residual amount (Bitcoins): used when an order is only partially satisfied by previous transactions;
\item limit price (see below), which in turn can be a real number;
\item time when the order was issued;
\item expiration time: if the order is not (fully) satisfied, it is removed from the book at this time.
\end{itemize}
		
The amount of each buy order depends on the amount of cash, $c_i(t)$, owned by $i$-th trader at time $t$, less the cash already committed to other pending buy orders still in the book. Let us call $c^b_i$ the available cash. The number of Bitcoins to buy, $b_a$ is given by eq. \ref{eq-buy}

\begin{equation}\label{eq-buy}
b_a = c^b_i p(t) \beta
\end{equation}

where $p(t)$ is the current price and $\beta$ is a random variable drawn from a lognormal distribution with average and standard deviation equal to $0.25$ and $0.2$, respectively. In the unlikely case that $\beta > 1$, $\beta$ is set equal to 1. In this way, a buyer posts on average orders equal to one fourth of her cash availability.

Similarly, the amount of each sell order depends on the number of Bitcoins, $b_i(t)$ owned by $i$-th trader at time $t$, less the Bitcoins already committed to other pending sell orders still in the book, overall called $b^s_i$. 
The number of Bitcoins to sell, $s_a$ is given by $s_a = b^s_i \beta$, where $\beta$ is a lognormal random variable as above.
Short selling is not allowed. 

The limit price models the price to which a trader desire to conclude his/her transaction. An order can also be issued with no limit
(market order), meaning that its originator wishes to perform the trade at the best price she can find. In this case, the limit 
price is set to zero. The probability of placing a market order, $P_{lim}$, is set at the beginning of the simulation 
and is equal to 0.2 for Random Traders and to 0.7 for Chartists.
This because, unlike Random Traders, if Chartists issue orders then they wish to perform the trade at the best available price,
to be able to gain by following the price trend.

Let us suppose that $i$-th trader issues a limit order to buy $a_i^b(t)$ Bitcoins at time $t$. 
Each buy order can be executed if the trading price is lower than, or equal to, the buy limit price $b_{i}$. 
In the case of a sell order of $a_i^s(t)$ Bitcoins, it can be executed if the trading price is higher than, or equal to, sell limit price $s_{i}$.
As said above, if the limit prices $b_{i}=0$ or $s_{i}=0$, then the orders can be always executed, provided there is a pending complementary order.

The buy and sell limit prices, $b_{i}$ and $s_{i}$, are given respectively by the following equations:

\begin{equation}\label{buyLimit}
b_{i}(t)=p(t)*N_i(\mu,\sigma_i)
\end{equation}

\begin{equation}\label{sellLimit}
s_{i}(t)=\frac{p(t)}{N_i(\mu,\sigma_i)}
\end{equation}

where
\begin{itemize}
\item $p(t)$ is the current Bitcoin price; 
\item $N_i(\mu,\sigma^{c}_i)$ is a random draw from a Gaussian distribution with average $\mu$ and standard deviation $\sigma_i$.
\end{itemize}

As you can see, the limit prices have a random component, modelling the fact that what traders "feel" is the right price to buy or to sell is not constant, and may vary for each single order. 
In the case of buy orders, we stipulate that a trader wishing to buy must offer a price that is, on average, slightly higher than the market price. We set this average to be 2\% higher: $\mu = 1.02$.

The value of $\sigma_i$ is proportional to the "volatility"  $\sigma(T_i)$ of the price $p(t)$ through the equation $\sigma_i=K\sigma(T_i)$, where $K$ is a constant and $\sigma(T_i)$ is the standard deviation of price absolute returns, calculated in the time window $T_i$ (this is the same approach of \cite{Raberto2001}). 

In the case of sell orders, the reasoning is dual. In this case, for symmetry, the limit price is divided by a random draw from the same Gaussian distribution $N_i(\mu,\sigma^{c}_i)$.

Each order issued can be satisfied only within a specific time interval. So, an expiration time is associated to each order. 
For Random Traders, the value of the expiration time is equal to the current time plus a number of days (time steps) drawn from a lognormal distribution with average and standard deviation equal to 3 and 1 days, respectively. In this way, most orders will expire within 4 days since they were posted.
Chartists, who act in a more dynamic way to follow the market trend, post orders whose expiration time is at the end of the same trading day.

Random traders issue buy or sell orders with the same probability, whereas Chartists issue buy orders when the price increases and sell orders when the price decreases.
In particular, they issue buy orders when the price relative variation in a time window $\tau_i$, different for each trader, is higher than a threshold $Th=0.01$, and issue sell orders if this variation is lower than $-Th$. 

\subsection{Price Clearing Mechanism}\label{sec:5a}

In the real crypto currency market, all Bitcoin transactions are recorded in the "Blockchain", which is a massive and transparent ledger of every Bitcoin transaction. The Blockchain is a mechanism that certifies the single transaction by using a timestamp server. This server generates computational proof of the chronological order of transactions to avoid the double-spending problem, and attacks of fraudsters.

In the Bitcoin system, clearing and settlement services are offered by several Bitcoin Exchanges. These exchanges operate quite
independently, but being the transaction price public and immediately available through inspecting the Blockchain, at least in principle they should work as a single exchange. 
These exchanges typically allow to post both limit orders and market orders. Some banks, such as South Africa's Standard Bank and Germany's Fidor Bank, are already offering price clearing services. 

Overall, we believe that a single, global Order Book mechanism can, at a first approximation, represent the Bitcoin market quite accurately. 
We implemented this mechanism as described below. It is similar to the book presented in \cite{Raberto2005}.

At every time step, the order book holds the list of all the orders received and
still to be executed. 
Buy orders are sorted in descending order with respect to the limit price $b_{i}$. Orders with the same limit price are sorted in ascending order with respect to the order issue time.
Sell orders are sorted in ascending order with respect to the limit price $s_{j}$. Orders with the same limit price are sorted in ascending order with respect to the order issue time.

At each simulation step, various new orders are inserted into the respective lists. 
As soon as a new order enters the book, the first buy order and the first sell order of the lists are inspected to verify if they match. 
If they match, a transaction occurs. The order with the smaller residual amount is fully executed, 
whereas the order with larger amount is only partially executed, and remains in the head of the list with residual amount reduced by the amount of the matching order. Clearly, if both orders have the same residual amount, they are both fully executed.

After the transaction, the next pair of orders at the head of the lists are checked for matching. If they match, they are executed, and so on until they do not match anymore. 
Hence, before the book can accept new orders, all the matching orders are satisfied.

A sell order of index $j$ matches a buy order of index $i$, and vice versa, only if $s_{j} \le b_{i}$, or if one of the two limit prices, or both, are equal to zero. 

As regards the price, $p_T$, to which the transaction is performed, the price formation mechanism follows the rules described below. Here, $p(t)$ denotes the current price:
\begin{itemize}
\item when one of the two orders has limit price equal to zero: 
\begin{itemize}
\item if $b_{i} > 0$, then $p_T=min(b_{i}, p(t))$,
\item if $s_{j} > 0$, then $p_T=max(s_{j}, p(t))$,
\end{itemize}
\item when both orders have limit price equal to zero, $p_T = p(t)$;
\item  when both orders have limit price higher than zero, $p_T = \frac{b_{i}+s_{j}}{2}$.
\end{itemize}

\section{Calibration of the Model}\label{sec:6}
The model described in the previous section was implemented in Smalltalk language. Before the simulation, it had to be calibrated in order to reproduce the real stylized facts in the Bitcoin market in the period between January 2012 and April 2014. The simulation period was thus set to 830 steps, a simulation step corresponding to to one day. We included also weekends and holidays, because the Bitcoin market is, by its very nature, accessible and working everyday.

\subsection{Trader Wealth Endowment}\label{sec:6.1}

At the beginning of the simulation, ($t=0$), each trader $i$ is endowed with an amount of cash $c_i(0)$ and an amount of Bitcoins $b_i(0)$. The distributions of cash and Bitcoins follow a power-law with exponent $\alpha$. In all the performed simulation we set $\alpha=1$, a value yielding the distribution known as Zipf's law \cite{Newman}. Table \ref{tab:initial} gives information about the actual quantities used to initialize traders' wealth.

To create this type of distribution we used the ranking property of the power-law 
\cite{Takayasu}. This means that if we order the wealth of agents in decreasing order, the cash/Bitcoins of the $i$th agent is $i^{-\alpha}$ the cash/Bitcoins of the first agent. 
We therefore artificially created a population of agents whose cash and number of Bitcoins at time step 0 follows the ranking property of the Pareto law.

In fact, we started with the total initial number of Bitcoins available, $B_T$, that is a known quantity. If the total number of traders is $N_t$ and the number of Bitcoins owned by them, $b_i$, follows a Pareto law with exponent $\alpha = 1$, it is well-known from Harmonic-series theory that $B_T = b_1 ln(N_t) + \gamma$, where $\gamma$ is the Euler-Mascheroni constant and $b_1$ is the number of Bitcoins owned by the richest trader. The number of Bitcoins owned by $i$-th trader is consequently: $\frac{b_1}{i}$.

A similar approach was followed to create the set of traders who entered the simulation at $t>0$, but in this case the traders were only endowed with cash. In this case, we had no specific data to calibrate the wealth of these traders. We stipulated that the cash, $c^s_1$, of the richest trader is about five times the cash owned by the richest initial trader, and that the exponent of the Pareto law is in this case $\alpha = 0.6$. In this way, the cash of the poorest traders amounts to about 2,600\$, whereas the total cash of these traders is about 8.7 million \$. This figure is slightly higher than the total cash of the initial traders ($C_T(0)=400,000$), multiplied by the ratio of the number of traders entering at $t>0$ over the number of initial traders. Overall, we performed various simulations varying these parameters, with no significant impact on the results.

The set of "new" traders are generated before the simulation starts. When new traders are needed to enter the simulation, they are chosen randomly in this set.

The wealth endowment of traders was made for all traders, both initial and new, irrespectively of being Random or Chartist ones. Then, they were randomly selected for being Random or Chartis traders with  probability  $P_R=0.7$ and $P_C=0.3$, respectively.

\subsection{Model Initialization}\label{sec:6.2}
We set the initial value of several key parameter of the model by computing the corresponding parameters using the Blockchain Web site. 
The main assumption we made is to size the artificial market at about 1/100 of the real market, to be able to manage the computational load of the simulation.
Table \ref{tab:initial} shows the parameter values and their computation assumptions in detail.

\begin{table}
\caption{Values of simulation parameters and the assumptions behind them.}
\label{tab:initial}       
\scalebox{0.7}{
\begin{tabular}{ccp{8cm}}
\hline\noalign{\smallskip}
Param.&Initial Value&Description and discussion\\
\noalign{\smallskip}\hline\noalign{\smallskip}
$N_t(0)$&100&Number of initial traders. Obtained dividing by 100 the number of average unique Bitcoin addresses as of January 2012 (which are about 10,000).\\
\hline
$B_T(0)$&80,000&Total initial number of Bitcoins. Obtained dividing by 100 the Bitcoins mined as of January 2012, that are approximately 8 million.\\
\hline
$p(0)$&5.0 \$&Initial price. The average price as of January 2012.\\
\hline
$C_T(0)$& 400,000 \$&Total initial cash. Obtained through formula: $C_T(0)=B_T(0) p(0)$. In fact, at equilibrium, the total value of Bitcoins is approximately equal to the total cash available to traders.\\
\hline
$N_t(t_{max})$&1,500&Total number of traders at the end of the simulation ($t=t_{max}=830$. Obtained dividing by 100 the number of average univocal Bitcoin addresses as of April 2014 (which are on average about 150,000).\\
\hline
$c^s_1$& 400,000 \$&Total initial cash of the richest trader entering the simulation at $t>0$.\\
\hline
$P_{lim}$&0.2 and 0.7 &Probability that an order has no limit price, for Random Traders and Chartists respectively. We stipulate that one fifth of all Random Traders' orders are market orders, and that a Chartists' order is a market order with a probability equal to 0.7.\\
\hline
$K$&$10^{-2}$&Constant used in the computation of $\sigma_i$ (see eqs. \ref{buyLimit} and \ref{sellLimit}). Tuned to yield a reasonable variation of $\sigma_i$ with price volatility.\\
\hline
$T_i$&10&Number of time steps used to compute the standard deviation of prices.\\
\noalign{\smallskip}\hline
\end{tabular}
}
\end{table}

\section{Simulation Results}\label{sec:7}
\subsection{Bitcoin Prices}\label{sec:7.1}

\begin{figure}
\centering
 \includegraphics[width=0.45\textwidth]{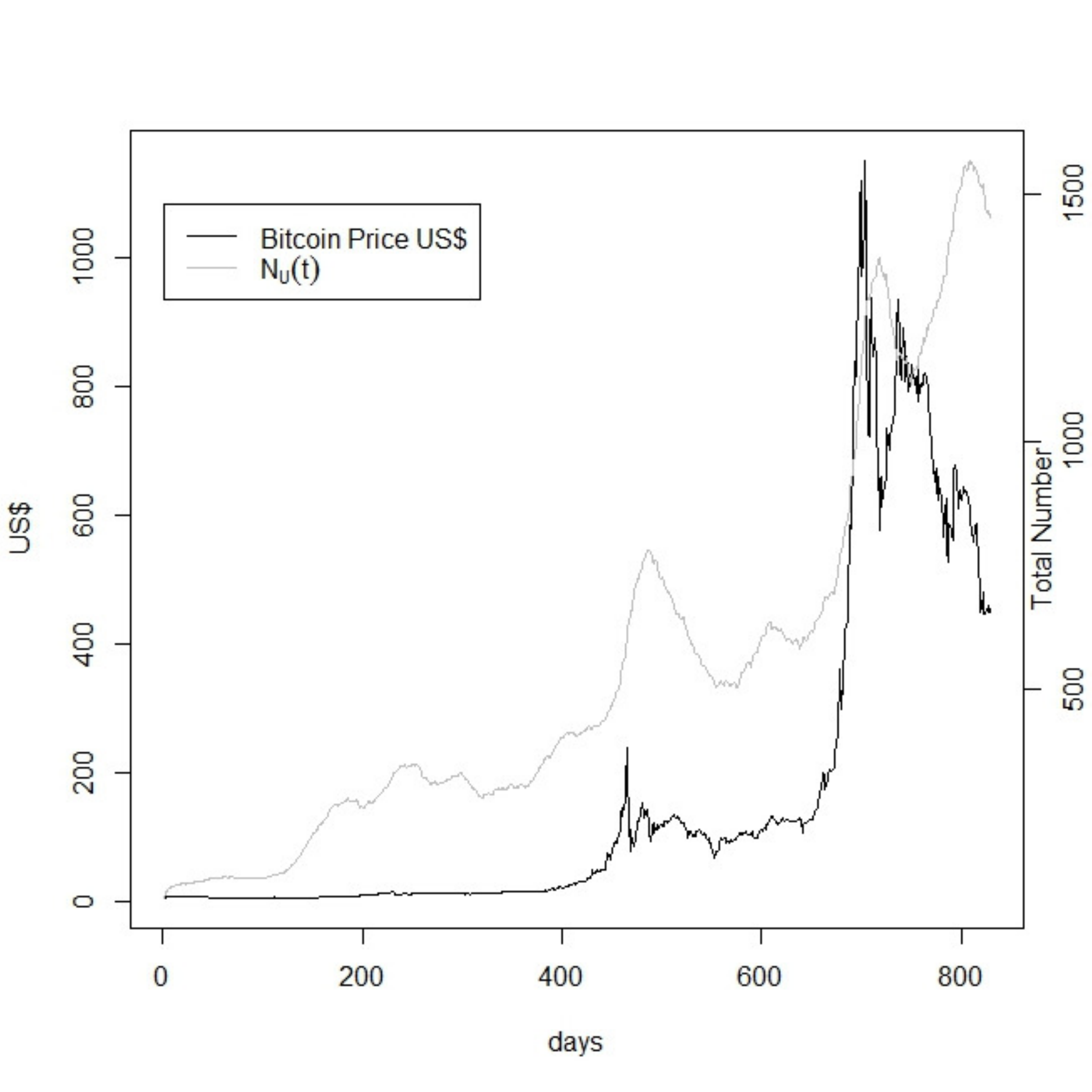}
\caption{Price of Bitcoins in US\$, and number of unique addresses of the Blockchain.}
\label{fig:realPrice}       
\end{figure}

We started studying the real Bitcoin price series between January 1, 2012 and April 10, 2014. The daily data were taken from the web site \textit{Blockchain.info}. 
Fig. \ref{fig:realPrice} shows the Bitcoin prices and the moving average of unique addresses of the Blockchain. It highlights a period in which the price trend is relatively constant, until about $450^{th}$ day. Then, a period of higher volatility follows between $450^{th}$ and $650^{th}$ day, followed by a period of high volatility after day 650, until the end of the considered interval.

The number of unique addresses tends to increase since the beginning of the considered period. Yet the price remains steady for a long time. However, the periods where the price increases most are also characterized by the largest increment of addresses. At the end of the considered interval, the price tends to decrease, whereas the number of adresses continues to increase.

It is well known that the price series encountered in financial markets typically exhibit some
statistical features, also known as "stylized facts" \cite{Pagan} \cite{Lux}. Among these, the three 
uni-variate properties which appear to be the most important and pervasive of price series, are 
(i) the unit-root property, (ii) the fat tail phenomenon, and (iii) the \emph{Volatility Clustering}.
We examined daily Bitcoin prices to assess whether also these prices exhibit these properties.

Regarding unit-root property, it amounts to being unable to reject the hypothesis that financial
prices follow a random walk. To this purpose, we applied the Augmented Dickey-Fuller test to the
series of Bitcoin daily prices we considered. 
The $\tau_3$ statistic for the null hypothesis $\gamma = 0$ is $-2.2$, and its
corresponding critical values at levels 1\%, 5\%, and 10\% with 830 observations
are $-3.96$, $-3.41$ and $-3.12$ respectively. At these levels
we can't reject the null hypothesis that $\gamma = 0$.


The second property is the fat-tail phenomenon: the distribution of absolute returns at weekly, daily
and higher frequencies displays a heavy tail with positive excess kurtosis.
In particular,the absolute return distribution in the tails can be approximated by a Pareto, or power-law, distribution:

\begin{equation}
P(x)= x^{-\alpha}
\end{equation}
where $\alpha $ is higher than zero and is known as the exponent or scaling parameter of power-law function.

The corresponding ccdf, i.e. the probability that the random variable is greater than a given value x, is:
\begin{equation}
P(X\ge x)\propto  x^{-(\alpha-1)}
\end{equation}

Given the expressions just described, it is simple to conclude that a power law is represented as a straight line in a 
log-log graph of both the probability density and the ccdf.

\begin{figure}
\centering
 \includegraphics[width=0.45\textwidth]{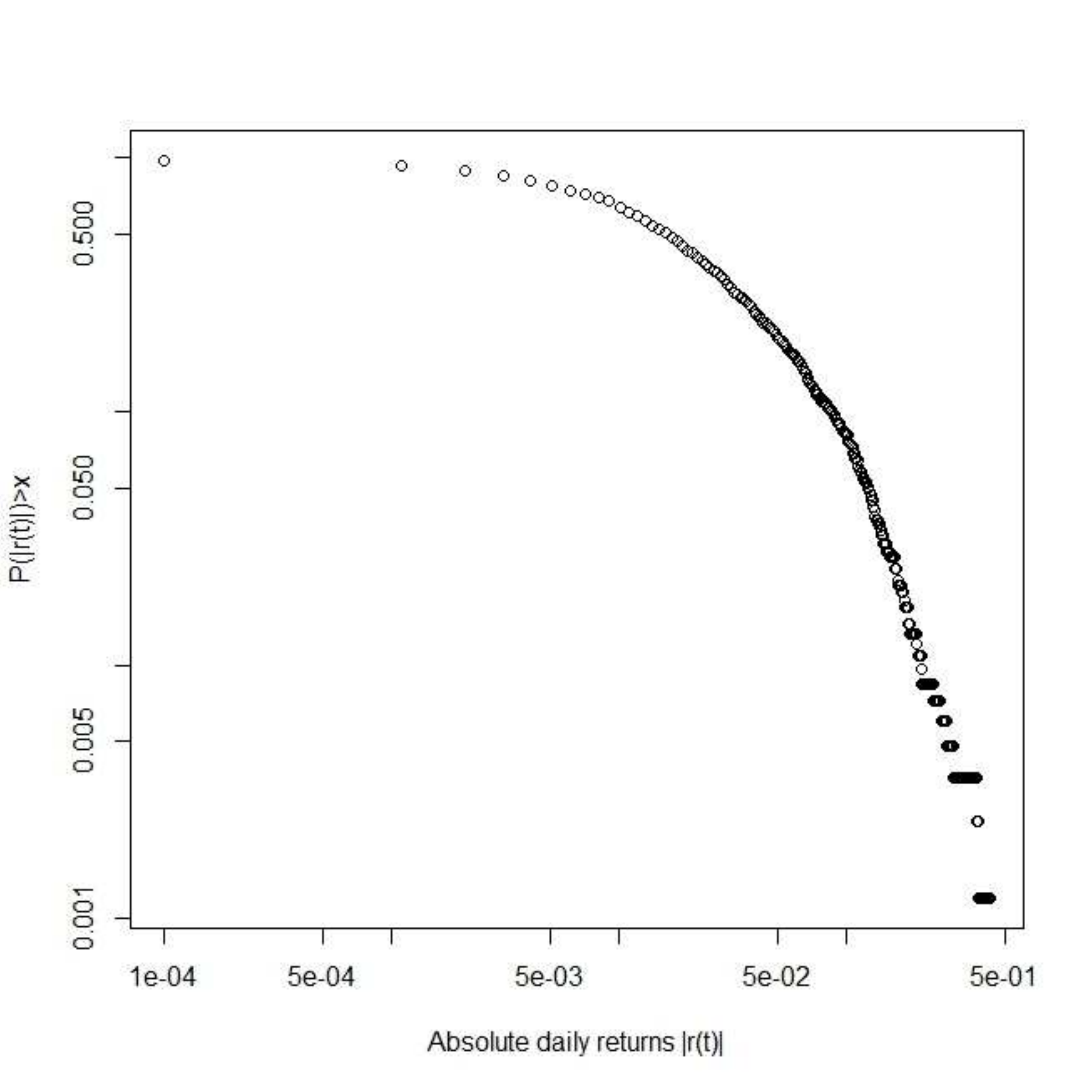}
\caption{Complementary cumulative distribution function of the price absolute returns.}
\label{fig:ccdfReal}       
\end{figure}

In figure \ref{fig:ccdfReal}, the log-log graph of the real absolute return ccdf is shown. The Pareto 
distribution holds for values of $x$ that are greater than a positive number $x_{min}$ -- in our case $x_{min} \simeq 10^{-1}$ -- 
and hence we can consider the ccdf of the real price absolute return as a power-law only in its right "tail", where
this curve is clearly a straight line. So, Bitcoin daily returns are not an exception to the "fat tail" stylized fact.

The third property is \emph{Volatility Clustering}: periods of quiescence and turbulence tend to cluster
together. This can be verified by the presence of highly significant autocorrelation in absolute or squared returns,
despite insignificant autocorrelation in raw returns. 

\begin{figure}[!ht]
\centering
\subfigure[]{
\includegraphics[width=0.45\textwidth]{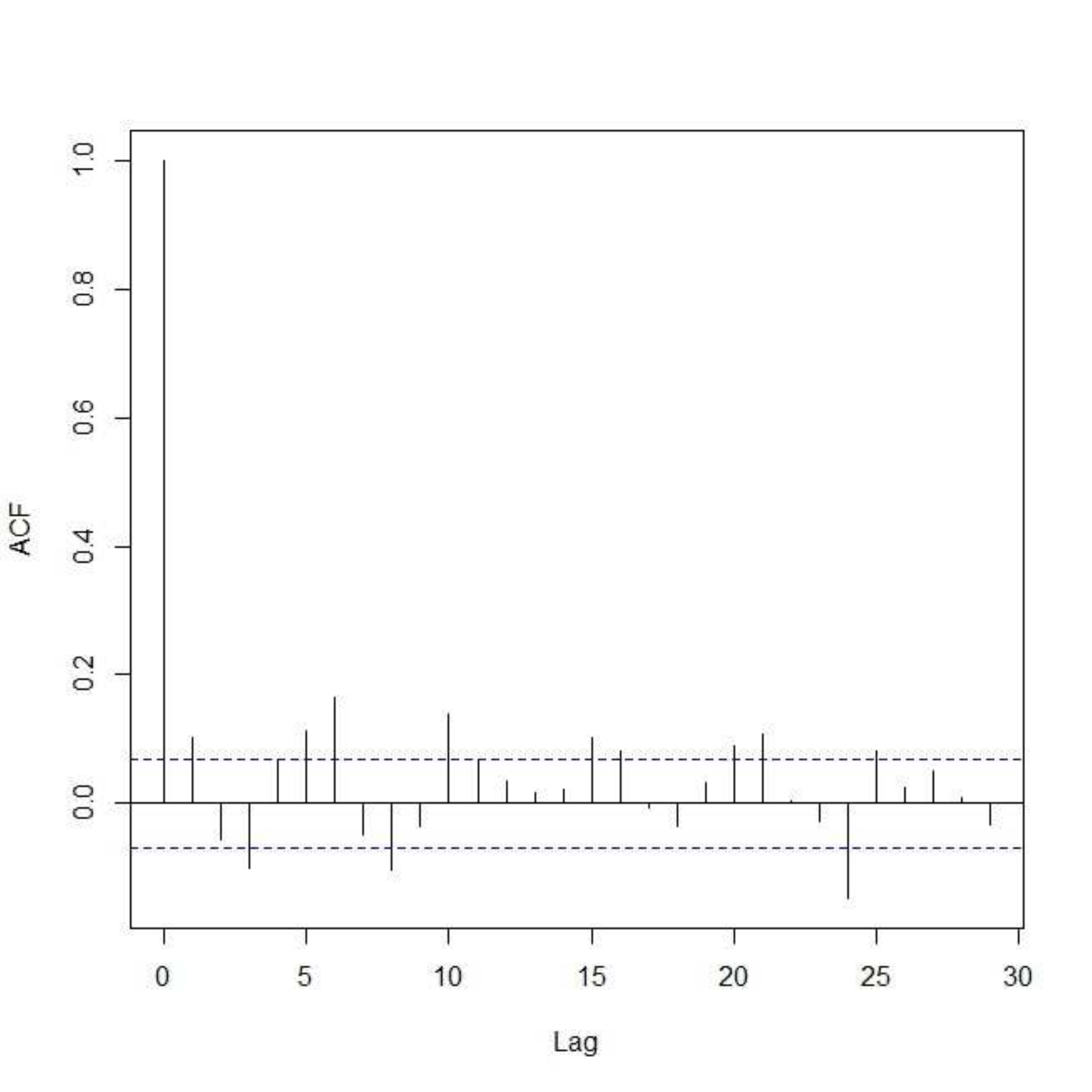}}
\hspace{7mm}
\subfigure[]{
\includegraphics[width=0.45\textwidth]{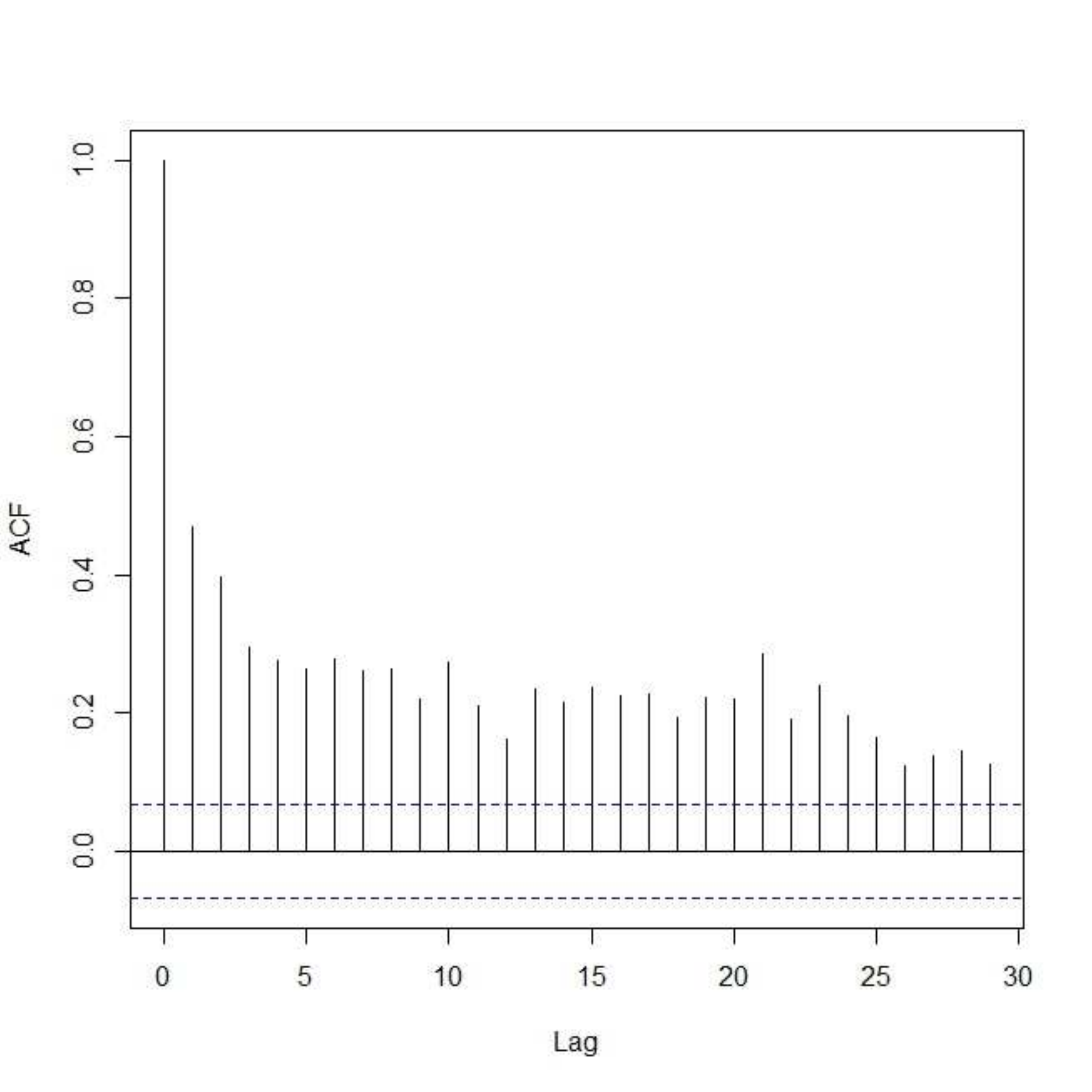}}
\caption{Autocorrelation of (a) raw returns, and (b) absolute returns of Bitcoin prices.  \label{fig:realAcf}}
\end{figure}

Fig. \ref{fig:realAcf} shows the autocorrelations of the real price returns and absolute returns, at different time lags. 
It is possible to note that the autocorrelation of raw returns (a) is often negative, and is anyway very close to zero, 
whereas the autocorrelation of absolute returns (b) slowly decreases for increasing lags, and that for most 
lag values is higher than 0.2. 
This behavior is typical of financial price return series, and confirms the presence of volatility clustering.

In conclusion, the Bitcoin price shows all the stylized facts of financial price series, as expected.

\subsection{Prices of the simulated model}\label{sec:7.2}

The model was run to study the main features which characterize the Bitcoin market and the traders who operate in it.

\begin{figure}
\centering
 \includegraphics[width=0.45\textwidth]{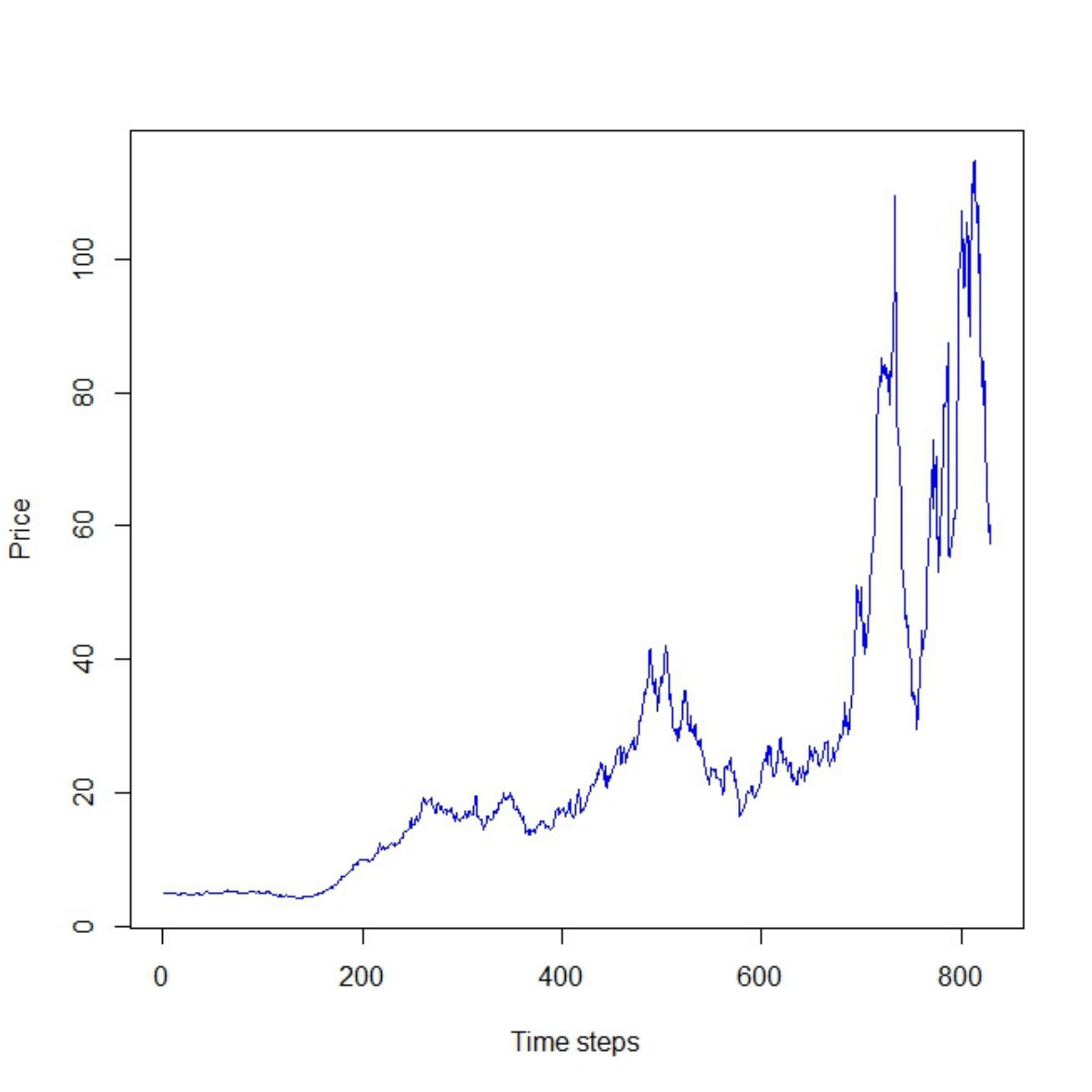}
\caption{Price of the Bitcoin in the simulated market.}
\label{fig:Price}       
\end{figure}

We performed several simulations using the parameter values calibrated as described in Section \ref{sec:6}.
The results of all simulations were consistent. We report in Fig. \ref{fig:Price} the Bitcoin price in one
typical simulation run. At a first sight, the behavior is quite similar to the real price (see Fig. \ref{fig:realPrice}).
It is possible to observe that, as in the case of the real price, at first the price keeps its value constant, and then, after about 200 simulation steps, it presents intervals characterized by high volatility, with price peaks of about 40\$, 100\$ and 110\$.
These values are much lower than real Bitcoin price spikes. We attribute it to the fact that the intrinsic mean reverting behavior 
of the simulated model prevents prices to spike as in the real market. All other statistical properties of prices and returns, 
however, are reproduced quite well in our model.

We performed the Augmented Dickey-Fuller test to the series of simulated Bitcoin daily prices. 
The results for one of the simulations are the followings.
The $\tau_3$ statistic for the null hypothesis $\gamma = 0$ is $-2.889$, and its
corresponding critical values at levels 1\%, 5\%, and 10\% with 830 observations
are $-3.96$, $-3.41$ and $-3.12$ respectively. At these levels, also for the simulated data
we can't reject the null hypothesis that $\gamma = 0$.
All simulations yield similar results.

\begin{figure}
\centering
\includegraphics[width=0.45\textwidth]{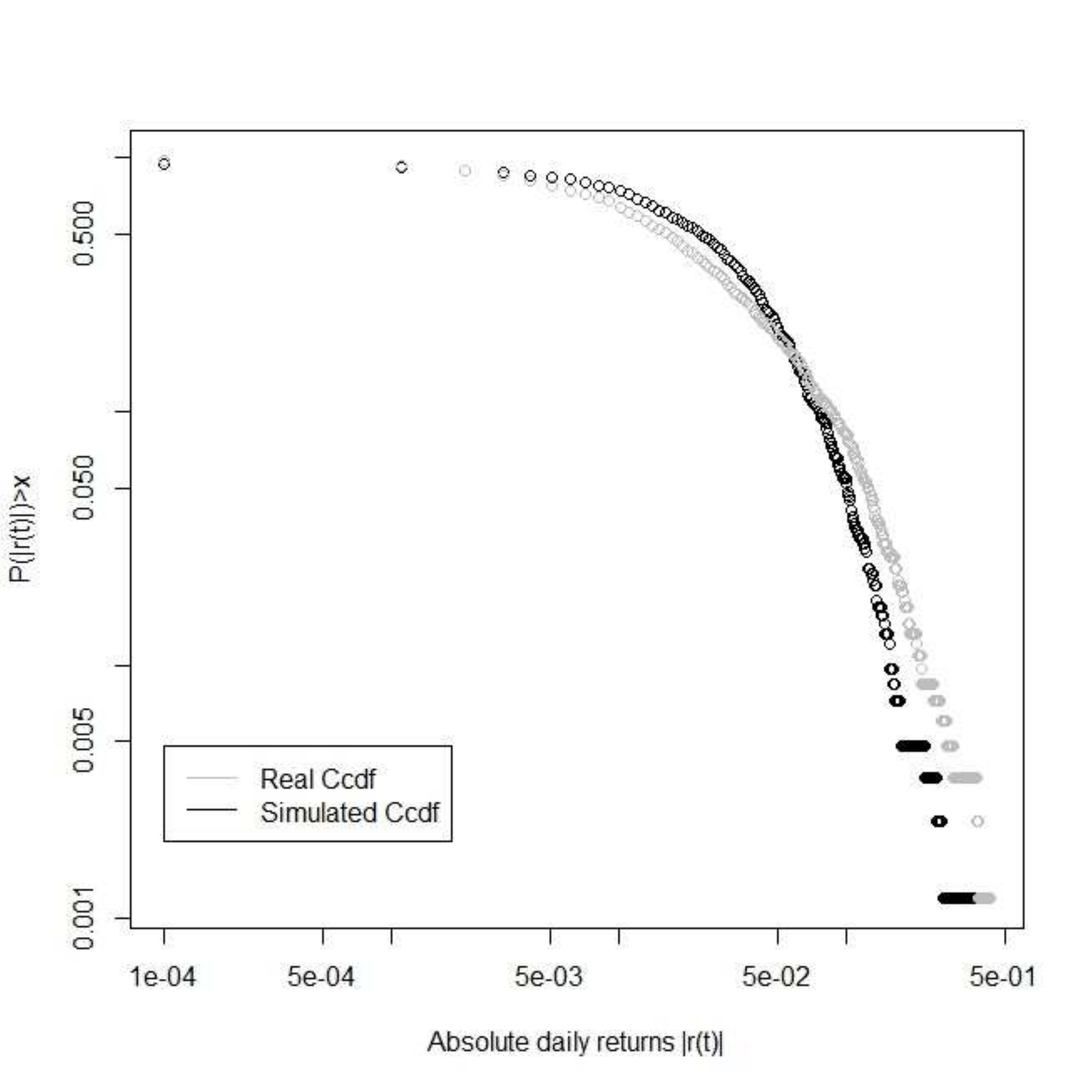}
\caption{Ccdf of the price absolute returns in a simulated Bitcoin market, compared with the real ccdf.}
\label{fig:ccdfSim}       
\end{figure}

The log-log plot of the ccdf of the price absolute returns in a simulated Bitcoin market is shown in Fig. \ref{fig:ccdfSim},
together with its real counterpart.
The linear behavior in the tail, that denotes a power-law, is patent. The two curves are quite similar,
though the real ccdf has a fat tail slightly broader.

\begin{figure}[!ht]
\centering
\subfigure[]{
\includegraphics[width=0.45\textwidth]{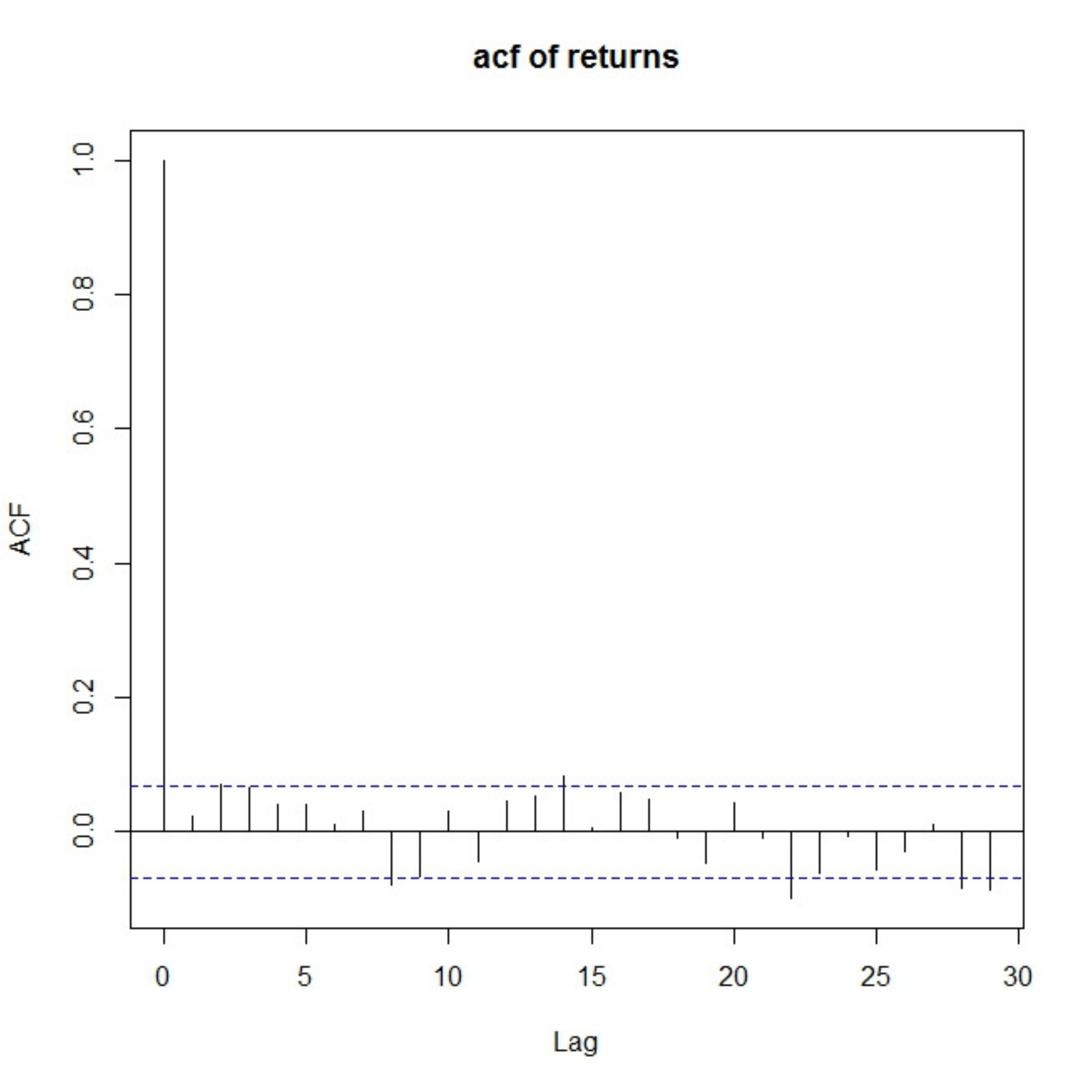}}
\hspace{7mm}
\subfigure[]{
\includegraphics[width=0.45\textwidth]{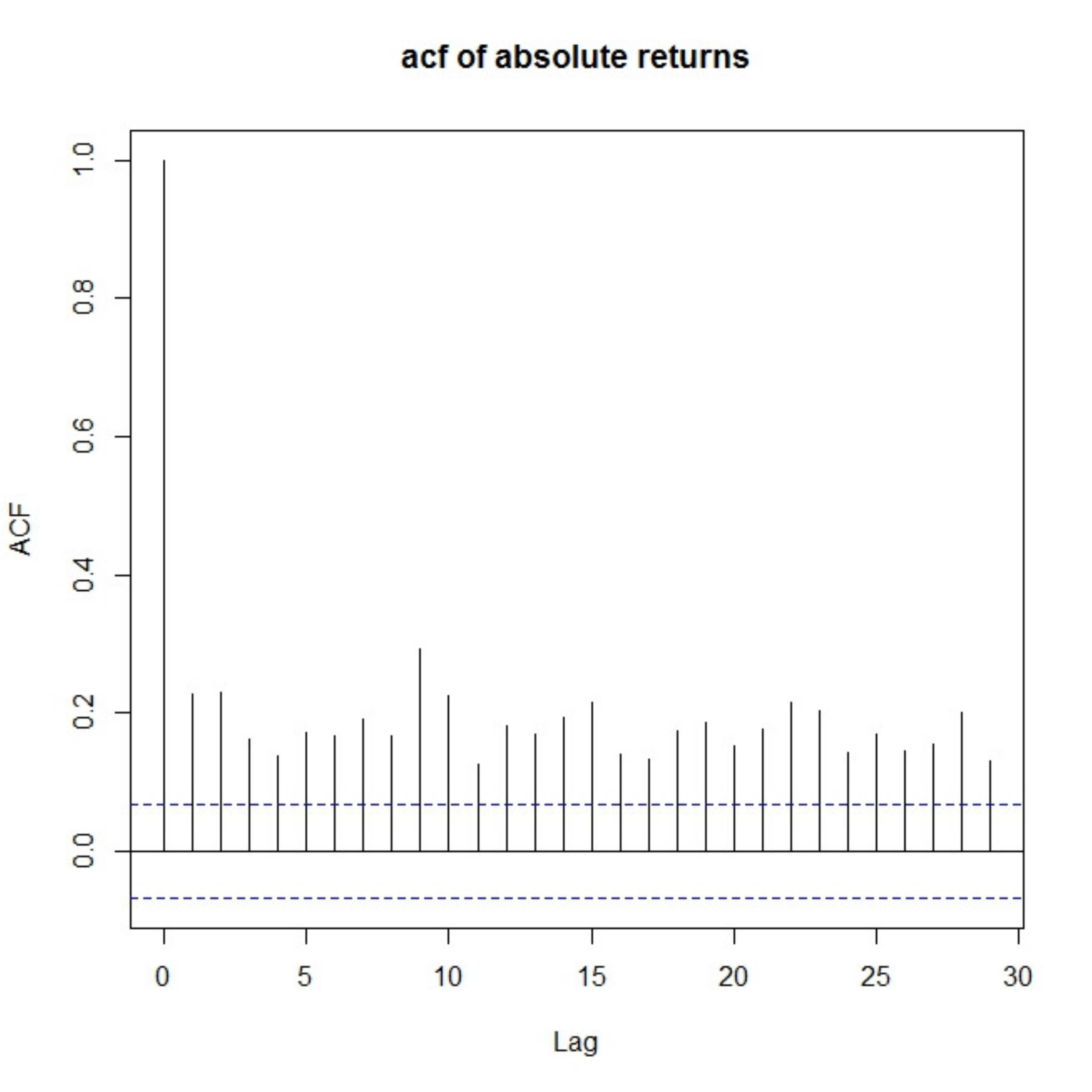}}
\caption{Autocorrelation of (a) raw returns, and (b) absolute returns of Bitcoin prices, in a simulated market. \label{fig:AcfCcdf}}
\end{figure}

The autocorrelations of the simulated price returns and absolute returns at different time lags 
are shown in Figs. \ref{fig:AcfCcdf} (a) and (b), respectively. 

The autocorrelation of raw returns (a) is much lower than that of absolute returns (b). Both are very
similar to those of real Bitcoin prices, shown in Fig. \ref{fig:realAcf}(a) and (b).
This confirms the presence of volatility clustering also for the simulated price series.

\subsection{Traders' statistics}\label{sec:7.3}

As described in the previous sections, each trader owns a specific amount of dollars and Bitcoins, so the amount of Bitcoins to be traded, depend on these quantities. The total wealth of $i$-th trader at time $t$, $w_i(t)$, is defined as the sum of her cash plus her Bitcoins multiplied by the current price, reported in eq. \ref{eq.wealth}. Of course, this quantity varies with the price.

\begin{equation}\label{eq.wealth}
w_i(t) = c_i(t) + b_i(t) p(t)
\end{equation}

\begin{figure}[!ht]
\centering
\subfigure[]{
\includegraphics[width=0.45\textwidth]{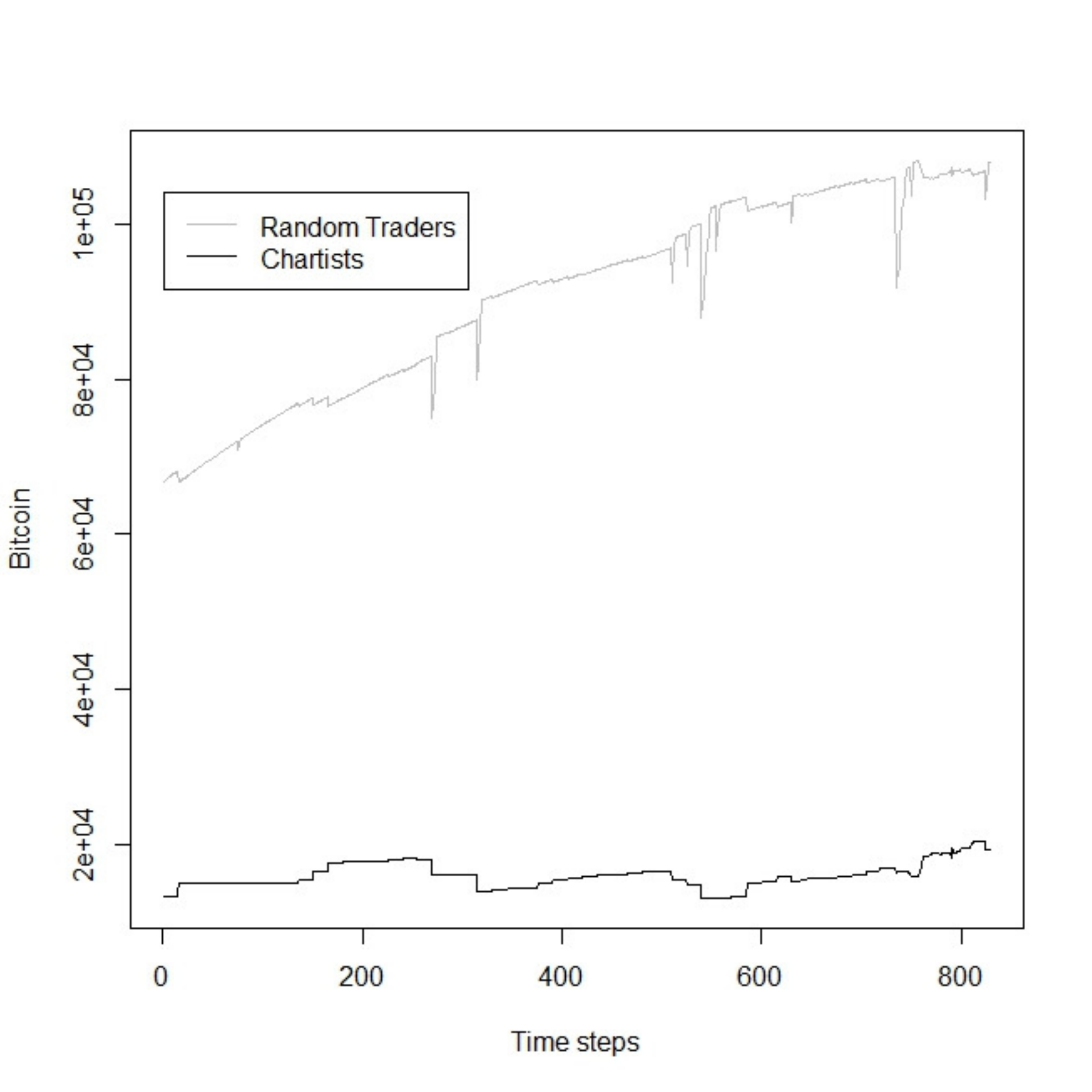}}
\hspace{7mm}
\subfigure[]{
\includegraphics[width=0.45\textwidth]{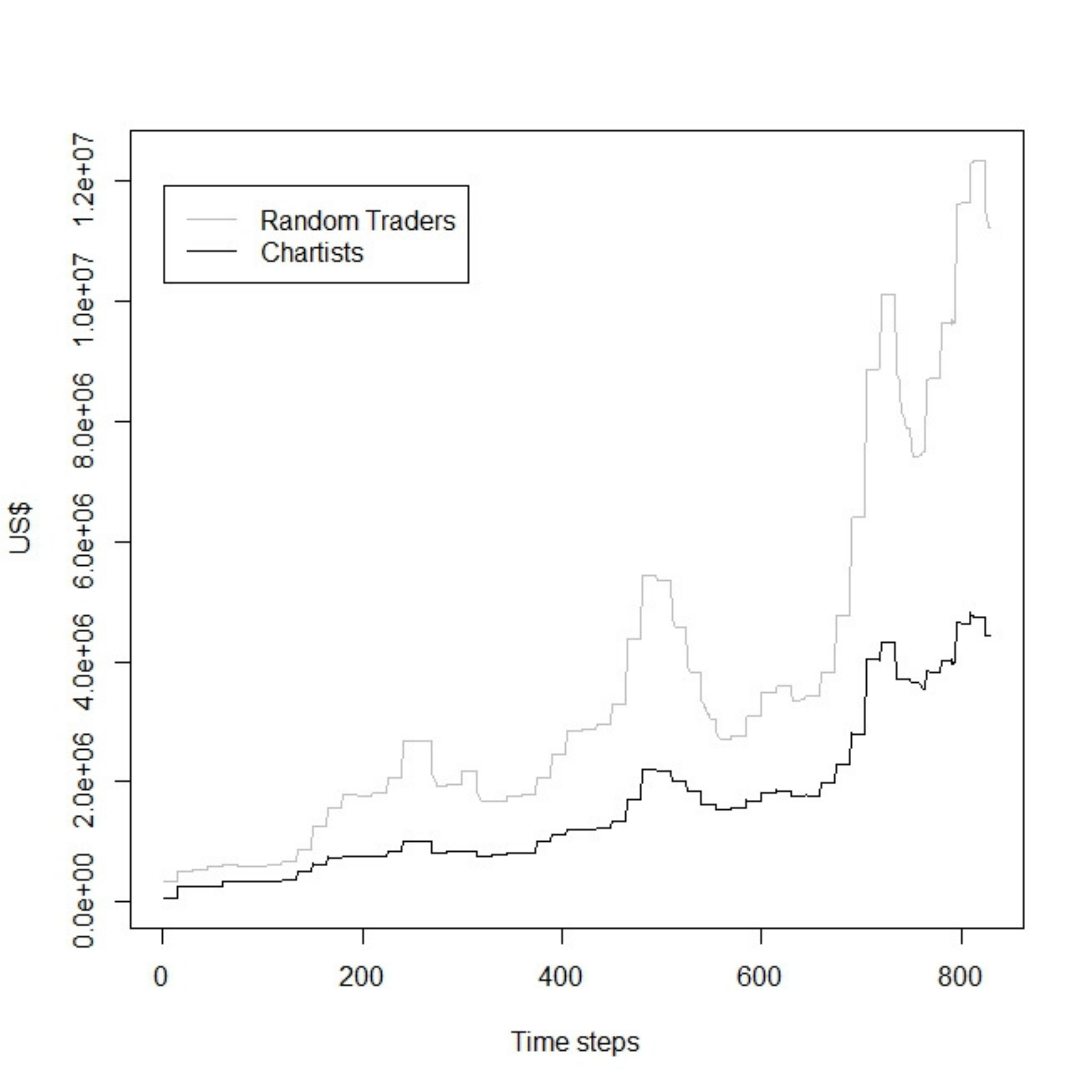}}
\caption{(a) Bitcoin amount and (b) cash amount for both trader populations during the simulation period. \label{fig:cash}}
\end{figure}

Some data about the distribution of the wealth among the different population of traders, the limit prices and the amount average of Bitcoin exchanged in each transaction are reported in the next figures.
Figs. \ref{fig:cash} (a) and (b) show the total amount of Bitcoins and cash owned by both populations, for all time steps. 
The "spikes" you see in Fig. \ref{fig:cash} (a) regarding Bitcoin endowment of Random traders are an artifact due to the removal of 
traders. When a trader is removed from the market, she sells all her Bitcoins through a market order. 
However, if there are no enough matching buy orders, these Bitcoins "disappear" from the market, until some trader buy them. 
Hence, the spikes. During the simulation shown there is no Chartist in a similar condition.

\begin{figure}[!ht]
\centering

\includegraphics[width=0.45\textwidth]{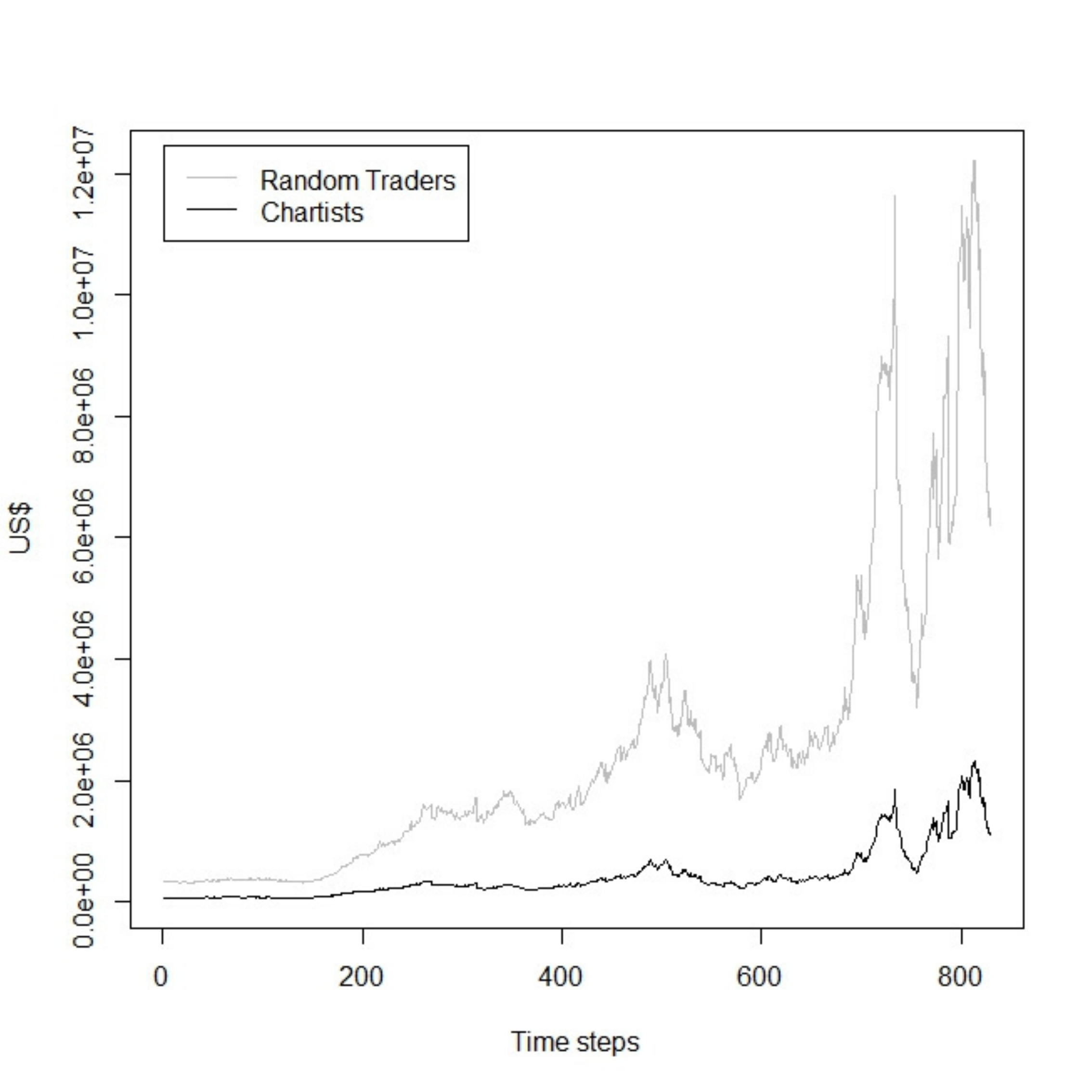}

\caption{Total wealth of both trader populations. \label{fig:cash1}}
\end{figure}

The sum of Bitcoins owned by Random and Chartist traders follows the trends shown in Fig. \ref{fig:basicData} (b). 
It is worth noting that the total amount of Bitcoins owned by Chartists looks fairly constant.

Fig. \ref{fig:cash1} shows the total wealth of both populations, as defined in eq. \ref{eq.wealth}. 
The total wealth is heavily influenced by price behavior. We also computed the ratio between the total
wealth of Random traders and that of Chartists, finding that it fluctuates around a constant value,
equal to the ratio of the total initial wealth of all Random traders over the total initial wealth of
Chartists. This ratio is on average equal the ratio of the number of Random over Chartist traders -- that is $0.7/0.3 \simeq 2.33$,
but it can differ substantially from the average, due to the Pareto distribution of traders' wealth. 
In fact, in different runs there can be relatively high differences between the total wealth of the various
kinds of traders, whereas the total wealth of all traders is constant.
The relative invariance of this ratio during the simulation means that the Chartist strategy is not winning
with respect to the Random strategy, but both are roughly equivalent.

\begin{figure}[!ht]
\centering
\subfigure[]{
\includegraphics[width=0.45\textwidth]{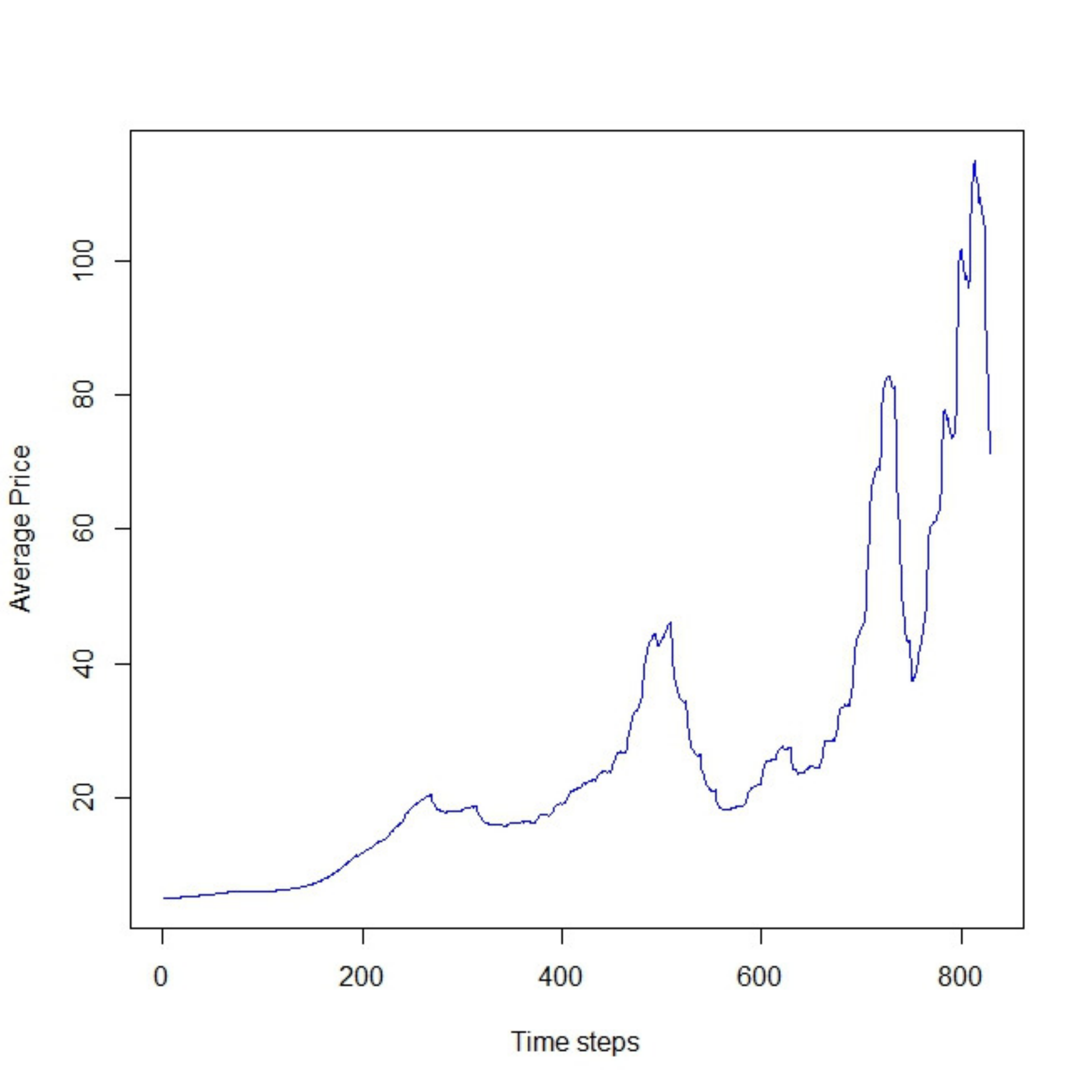}}
\hspace{7mm}
\subfigure[]{
\includegraphics[width=0.45\textwidth]{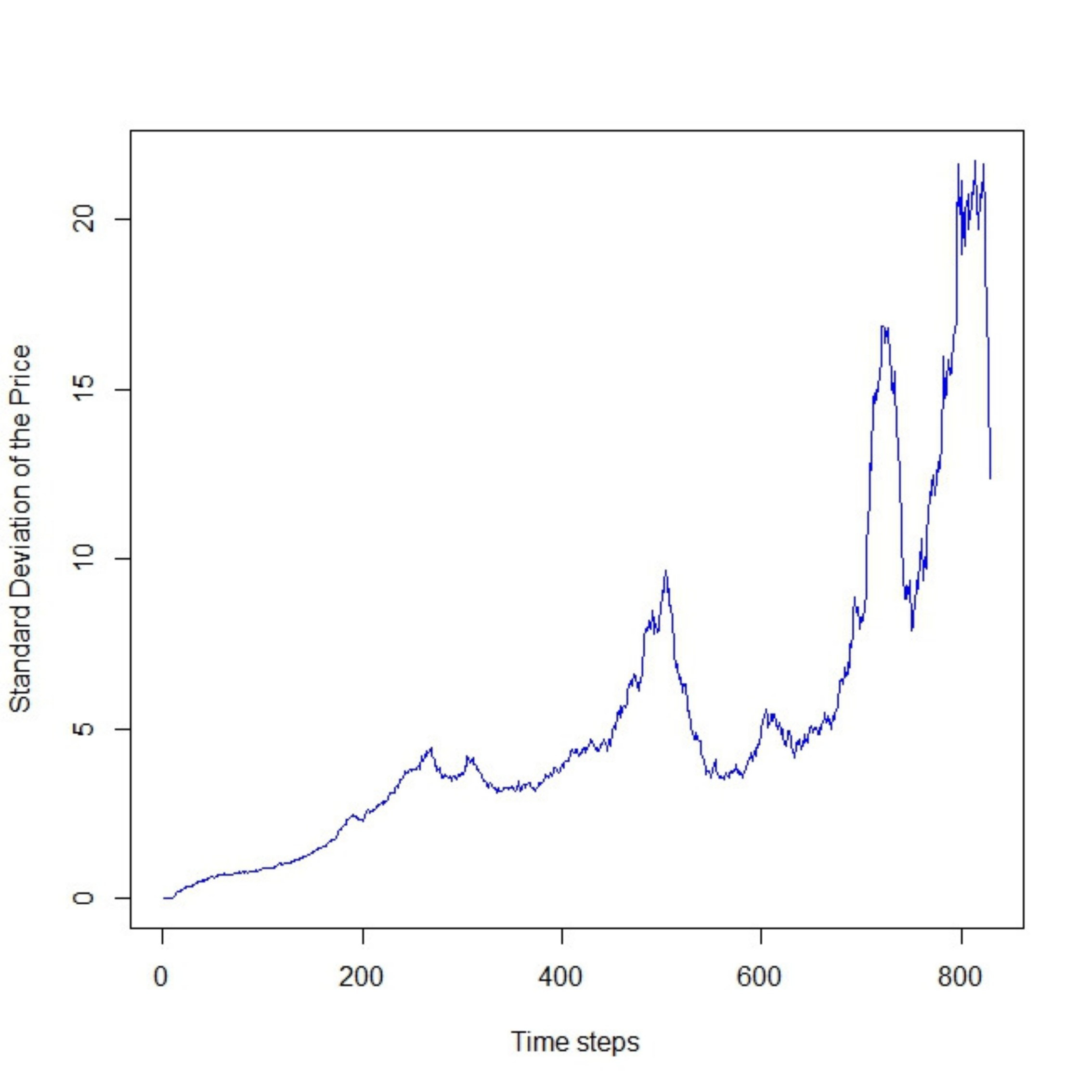}}
\caption{(a) Average Price  and (b) standard deviation computed on the 100 Monte Carlo simulations performed. \label{fig:averagePrice}}
\end{figure}

\begin{figure}[!ht]
\centering
\subfigure[]{
\includegraphics[width=0.45\textwidth]{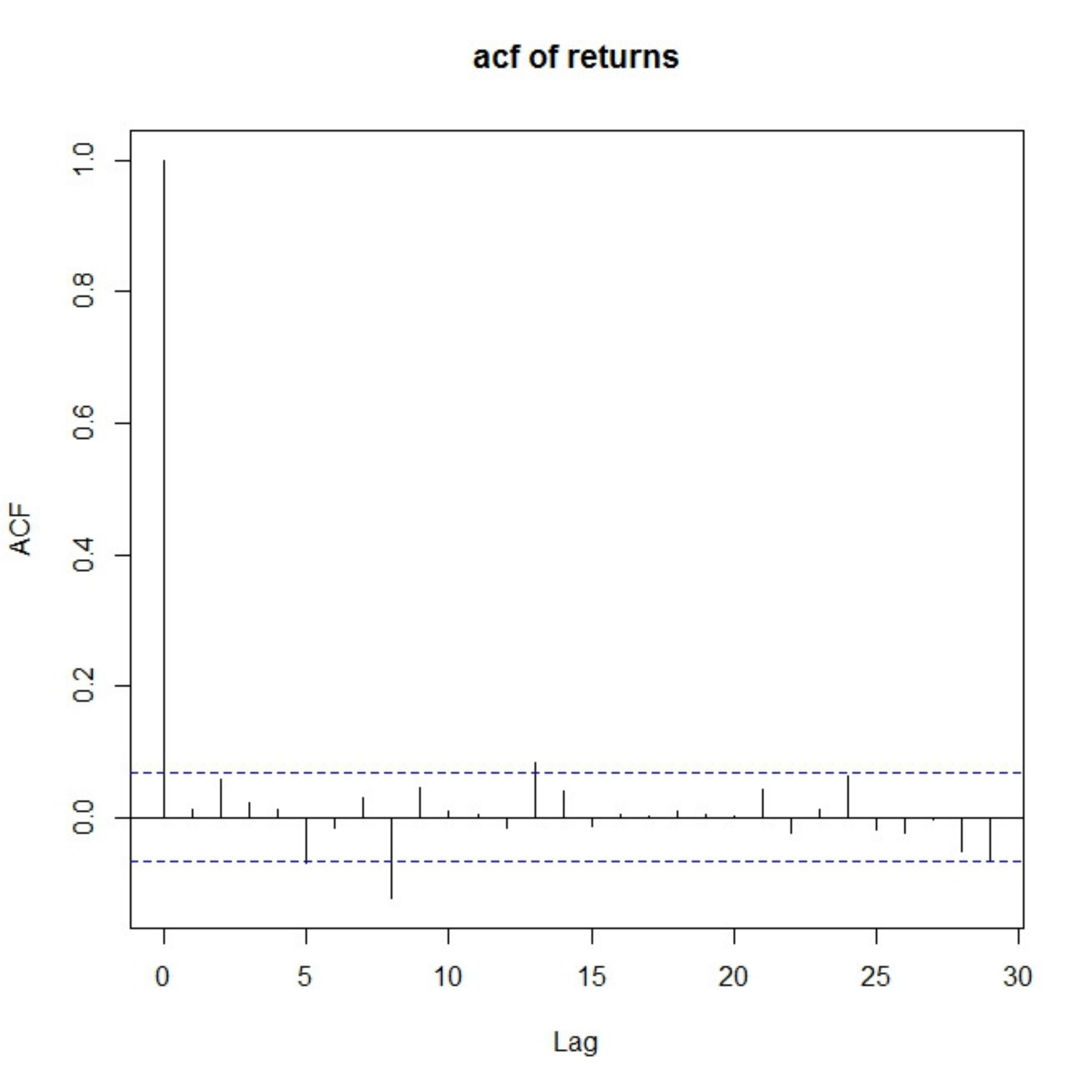}}
\hspace{7mm}
\subfigure[]{
\includegraphics[width=0.45\textwidth]{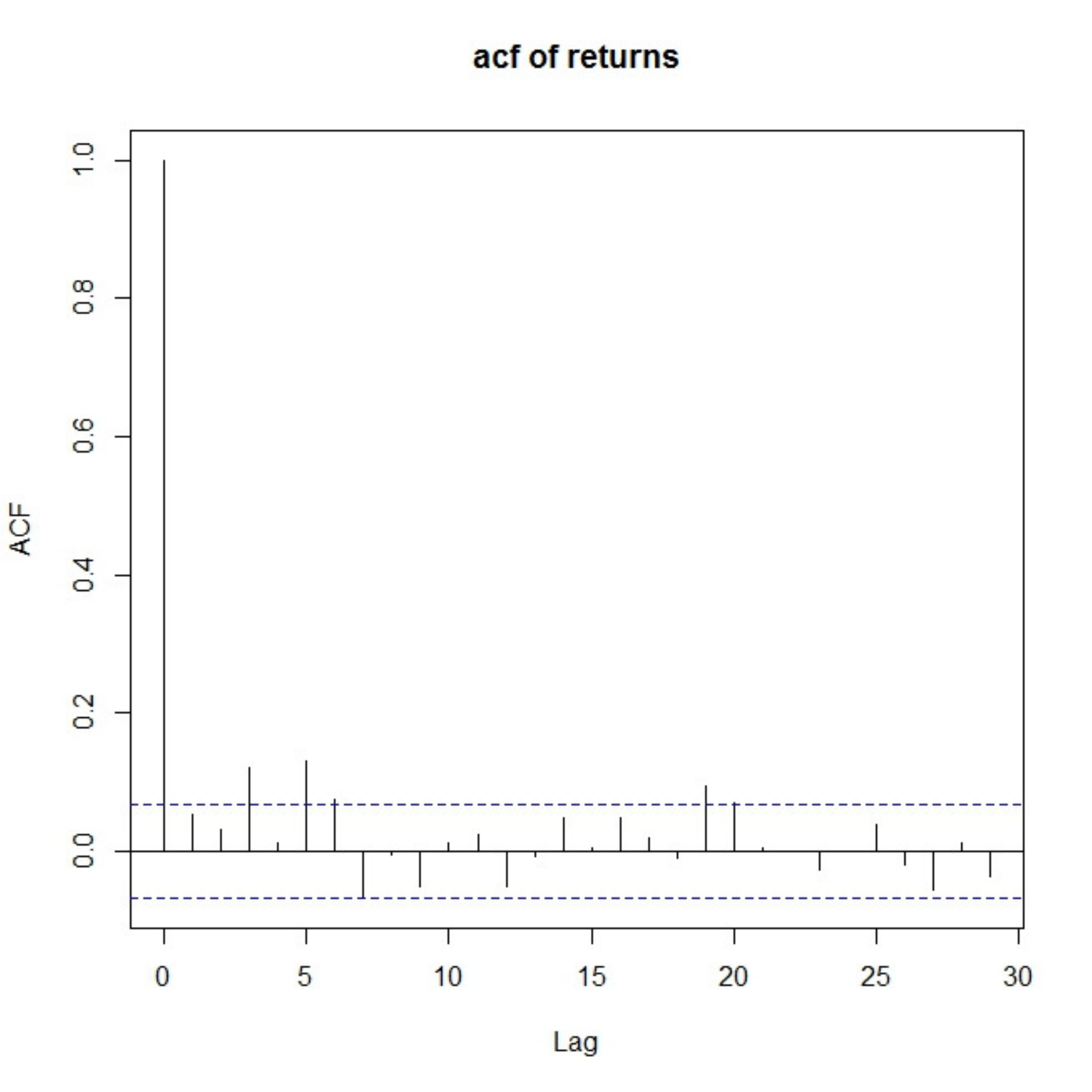}}
\caption{Examples of autocorrelations of the returns computed on two different Monte Carlo simulations. \label{fig:acf}}
\end{figure}

\subsection{Robustness analysis}

In order to assess the robustness of our model and the validity of our statistical analysis, we
repeated 100 simulations with the same initial conditions, but different seeds of the random number generator. 

Figs. \ref{fig:averagePrice} (a) and (b) report the average and the standard deviation of the simulated price, taken on all 100
simulations. The average price follows quite closely the number of traders present in the market, reported in Fig. \ref{fig:basicData}.
This is reasonable because, given the slowly varying amount of Bitcoins, the price should be roughly proportional to the 
cash available in the market to buy them, which is in turn proportional to the traders number.
The standard deviation of prices is roughly equal to $1/4$ of the average, meaning that the 100 simulations 
are quite consistent as regards price behavior.

Figs. \ref{fig:acf} (a) and (b) show the raw return autocorrelation functions (ACF) of a couple of simulations extracted from the 100 performed.
Both ACFs show the same behavior as reported in Fig. \ref{fig:AcfCcdf} (a). All 100 ACFs have similar behaviors.
Also the ACF of absolute returns show a behavior similar to that reported in Fig. \ref{fig:AcfCcdf} (b) for all the considered cases.

\begin{figure}[!ht]
\centering
\includegraphics[width=0.45\textwidth]{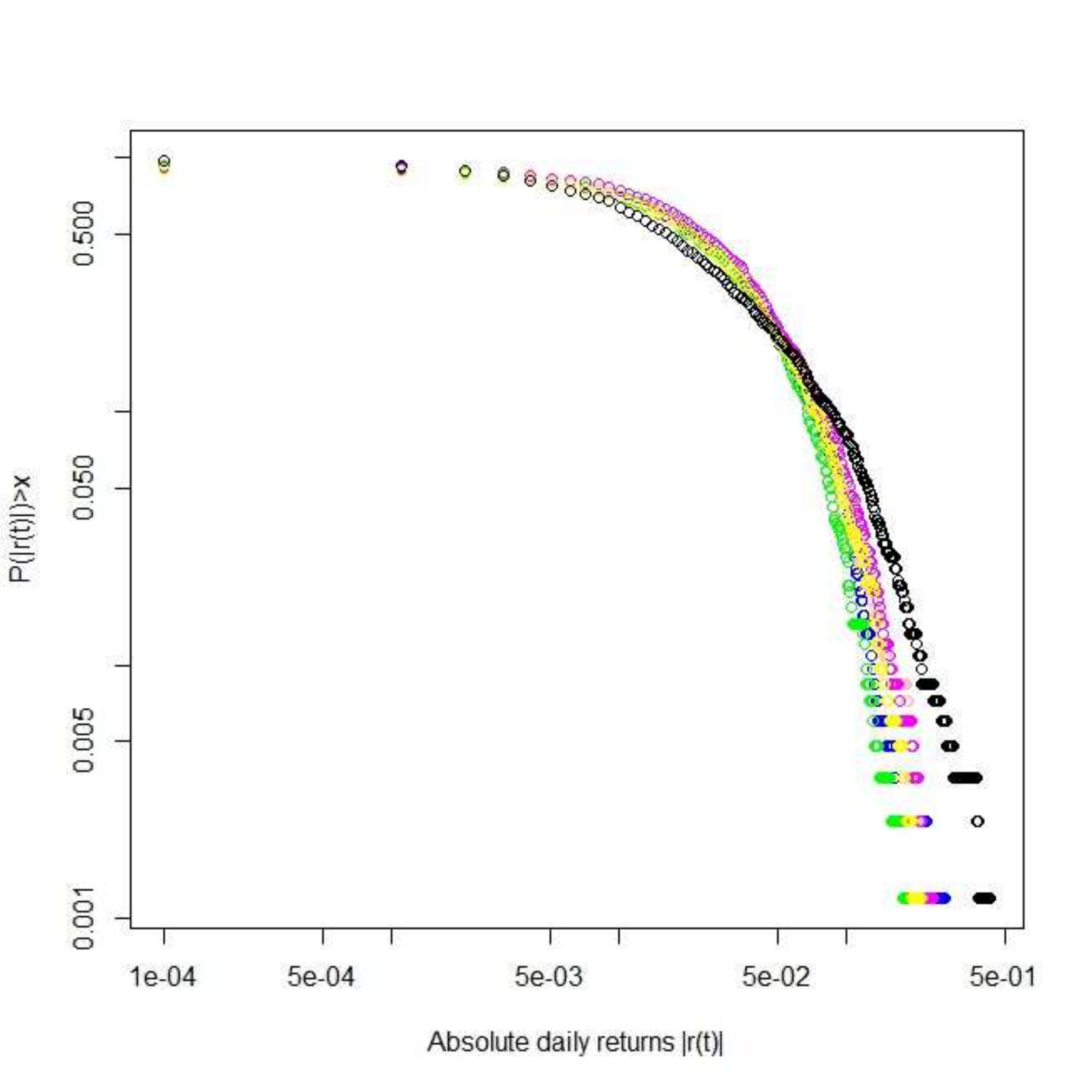}
\caption{(a) Complementary cumulative distribution function of absolute returns of real prices (black circles),
and ccdf of ten simulated cases taken at random (colored circles). \label{fig:acfSomeCcdf}}
\end{figure}

Figs. \ref{fig:acfSomeCcdf} show the cccdf of the price absolute returns in the real case and for ten simulations 
extracted from the 100 simulations quoted. Also in this case, the power-law behavior is patent in all reported simulations,
though with a greater slope than in the real case.

Overall, all performed simulations presented a consistent behavior, with no significant variations of statistical 
properties of the Bitcoin prices and of traders' endowment of Bitcoins and cash.

\section{Conclusions} \label{sec:8}
In this paper, we present an heterogenous agent model of the Bitcoin market, accurately modeling many of the characteristics
of the real market. Namely, the model includes different trading strategies, an initial distribution of wealth following
Pareto law, a realistic trading and price clearing mechanism based on an order book, the increase with time of the total 
number of Bitcoins due to mining, the arrival of new traders interested in Bitcoins.  
The model is simulated and its outputs -- especially the Bitcoin prices -- are analysed and compared to real prices.

The main result of the model, besides being to our knowledge the first model of a cryptocurrency market following the
artificial financial market approach, is the fact that some key stylized facts of Bitcoin real price series
are very well reproduced. 
The computational experiments performed produce price series for which we are unable to reject the hypothesis that they
follow a random walk. 
The autocorrelation of raw returns is very low for all time lags, whereas the autocorrelation of absolute returns 
is much higher, confirming the presence of volatility clustering.
Also, the ccdf of the absolute returns exhibit a power-law behavior in its tail, like that of real absolute returns.

Note that the results obtained are quite sensitive to the traders' behavior. 
We found that the presence of different traders' populations, and hence the trading between Random traders and Chartists,
is essential to reproduce the autocorrelation and the ccdf of the returns of the Bitcoin price. 
In particular, the Chartists' behaviour is essential to reproduce autocorrelations of the returns that confirm periods of quiescence and turbulence in the simulated Bitcoin price.

Future research will explore more traders' behaviors, a more detailed mechanism for describing mining, the
interplay between different cryptocurrencies through explicit network effects among traders. 

%
%
%
%

\end{document}